\documentclass[conference]{IEEEtran}

\usepackage{multirow}
\usepackage[table]{xcolor}
\definecolor{lightblue}{rgb}{0.93,0.95,1.0}

\usepackage{mathtools}

\usepackage{colortbl}

\usepackage{mdwmath}
\usepackage{algorithm}
\usepackage{array}

\usepackage{environ}
\usepackage[noend]{algpseudocode}
\usepackage{booktabs}
\usepackage{cite}

\usepackage[pdftex]{graphicx}

\usepackage{url}

\usepackage{stfloats}

\usepackage{epsfig}

\usepackage{tabularx}
\usepackage{rotating}
\usepackage{times}
\usepackage{comment}
\usepackage{amssymb}
\usepackage{amsthm}
\usepackage{pifont}

\usepackage{footnote}
\usepackage[bookmarks=false,colorlinks=true,citecolor=blue,linkcolor=red,urlcolor=blue]{hyperref}
\usepackage{xspace} 
\usepackage{setspace}
\usepackage{bm}
\usepackage[export]{adjustbox}
\usepackage{stfloats} 
\usepackage{microtype} 
\usepackage{diagbox}
\usepackage{listings}

\usepackage[caption=false,font=footnotesize]{subfig}

\usepackage{soul}

\microtypecontext{spacing=nonfrench}

\newcommand{\ie}{\emph{i.e., }}
\newcommand{\eg}{\emph{e.g., }}

\theoremstyle{definition}

\newcommand{\cmark}{\ding{51}}%
\newcommand{\xmark}{\ding{55}}%

\definecolor{Gray}{gray}{0.85}
\definecolor{darkgray}{rgb}{.4,.4,.4}
\definecolor{lightgray}{HTML}{B3B3B3}
\definecolor{myblue}{HTML}{A9C4EB}
\definecolor{lightred}{HTML}{FF9999}
\definecolor{lightorange}{HTML}{FFCC99}

\newcolumntype{?}[1]{!{\vrule width #1}}

\def\sys{\textsc{Trex}\xspace}

\newcommand{\etal}{\textit{et al.~}}

\DeclareMathOperator*{\argmin}{arg\,min}

\begin{document}
\title{\sys: Learning Execution Semantics from\\ Micro-Traces for Binary Similarity}

\makeatletter
\newcommand{\linebreakand}{%
  \end{@IEEEauthorhalign}
  \hfill\mbox{}\par
  \mbox{}\hfill\begin{@IEEEauthorhalign}
}
\makeatother

\author{\IEEEauthorblockN{Kexin Pei}
\IEEEauthorblockA{Columbia University\\
kpei@cs.columbia.edu}
\and
\IEEEauthorblockN{Zhou Xuan}
\IEEEauthorblockA{University of California, Riverside\\
zxuan004@ucr.edu}
\linebreakand
\IEEEauthorblockN{Junfeng Yang}
\IEEEauthorblockA{Columbia University\\
junfeng@cs.columbia.edu}
\and
\IEEEauthorblockN{Suman Jana}
\IEEEauthorblockA{Columbia University\\
suman@cs.columbia.edu}
\and
\IEEEauthorblockN{Baishakhi Ray}
\IEEEauthorblockA{Columbia University\\
rayb@cs.columbia.edu}}

\date{}

\maketitle

\begin{abstract}

Detecting semantically similar functions -- a crucial analysis capability with broad real-world security usages including vulnerability detection, malware lineage, and forensics -- requires understanding function behaviors and intentions. However, this task is challenging as semantically similar functions can be implemented differently, run on different architectures, and compiled with diverse compiler optimizations or obfuscations. Most existing approaches match functions based on syntactic features without understanding the functions' execution semantics.

We present \sys, a transfer-learning-based framework, to automate learning execution semantics explicitly from functions' micro-traces (a form of under-constrained dynamic traces) and transfer the learned knowledge to match semantically similar functions. While such micro-traces are known to be too imprecise to be directly used to detect semantic similarity, our key insight is that these traces can be used to teach an ML model the execution semantics of different sequences of instructions. We thus design an unsupervised pretraining task, which trains the model to learn execution semantics from the functions' micro-traces without any manual labeling or feature engineering effort. We then develop a novel neural architecture, hierarchical Transformer, which can learn execution semantics from micro-traces during the pretraining phase. 
Finally, we finetune the pretrained model to match semantically similar functions.

We evaluate \sys on 1,472,066 function binaries from 13 popular software projects. These functions are from different architectures (x86, x64, ARM, and MIPS) and compiled with 4 optimizations (\texttt{O0}-\texttt{O3}) and 5 obfuscations. \sys outperforms the state-of-the-art systems by 7.8\%, 7.2\%, and 14.3\% in cross-architecture, optimization, and obfuscation function matching, respectively, while running 8$\times$ faster. Our ablation studies show that the pretraining task significantly boosts the function matching performance, underscoring the importance of learning execution semantics. Moreover, our extensive case studies demonstrate the practical use-cases of \sys\xspace -- on 180 real-world firmware images with their latest version, \sys uncovers 16 vulnerabilities that have not been disclosed by any previous studies. 
We release the code and dataset of \sys at \url{https://github.com/CUMLSec/trex}.

\end{abstract}

\section{Introduction}
\label{sec:intro}

Semantic function similarity, which quantifies the behavioral similarity between two functions, is a fundamental program analysis capability with a broad spectrum of real-world security usages, such as vulnerability detection~\cite{brumley2008automatic}, exploit generation~\cite{avgerinos2014automatic}, tracing malware lineage~\cite{jang2013towards, bayer2009scalable}, and forensics~\cite{luo2014semantics}.
For example, OWASP lists ``using components with known vulnerabilities'' as one of the top-10 application security risks in 2020~\cite{OWASP}.
Therefore, identifying similar vulnerable functions in massive software projects can save significant manual effort.

When matching semantically similar functions for security-critical applications (\eg vulnerability discovery), we often have to deal with software at \emph{binary level}, such as commercial off-the-shelf products (\ie firmware images) and legacy programs.
However, this task is challenging, as the functions' high-level information (\eg data structure definitions) are removed during the compilation process. 
Establishing semantic similarity gets even harder when the functions are compiled to run on different instruction set architectures with various compiler optimizations or obfuscated with simple transformations.

Recently, Machine Learning (ML) based approaches have shown promise in tackling these challenges~\cite{xu2017neural, massarelli2019safe, ding2019asm2vec} by learning robust features that can identify similar function binaries across different architectures, compiler optimizations, or even some types of obfuscation. Specifically, ML models learn function representations (\ie embeddings) from function binaries and use the distance between the embeddings of two functions to compute their similarity. The smaller the distance, the more similar the functions are to each other.  Such approaches have achieved state-of-the-art results~\cite{xu2017neural, massarelli2019safe, ding2019asm2vec}, outperforming the traditional signature-based methods~\cite{bindiff} using hand-crafted features (\eg number of basic blocks).  Such embedding distance-based strategy is particularly appealing for large-scale function matching---taking only around 0.1 seconds searching over one million functions~\cite{feng2016scalable}.

\vspace{.1cm}\noindent\textbf{Execution semantics.}
Despite the impressive progress, it remains challenging for these approaches to match semantically similar functions with disparate syntax and structure~\cite{mckee2019software}. 
An inherent cause is that the code semantics is characterized by \emph{its execution effects}. 
However, all existing learning-based approaches are \emph{agnostic to program execution semantics}, training only on the static code. 
Such a setting can easily lead a model into matching simple patterns, limiting their accuracy when such spurious patterns are absent or changed~\cite{payer2014similarity, aghakhani2020malware}.

For instance, consider the following pair of x86 instructions: \texttt{mov eax,2;lea ecx,[eax+4]} are semantically equivalent to \texttt{mov eax,2;lea ecx,[eax+eax*2]}. 
An ML model focusing on syntactic features might pick common substrings (both sequences share the tokens \texttt{mov}, \texttt{eax}, \texttt{lea}, \texttt{ecx}) to establish their similarity, which does not encode the key reason of the semantic equivalence. 
Without grasping the approximate execution semantics, an ML model can easily learn such spurious patterns without understanding the inherent cause of the equivalence: \texttt{[eax+eax*2]} computes the same exact address as \texttt{[eax+4]} when \texttt{eax} is 2.

\vspace{.1cm}\noindent\textbf{Limitations of existing dynamic approaches.}
Existing dynamic approaches try to avoid the issues described above by directly comparing the dynamic behaviors of functions to determine similarity. 
As finding program inputs reaching the target functions is an extremely challenging and time-consuming task, the prior works perform under-constrained dynamic execution by initializing the function input states (\eg registers, memory) with random values and executing the target functions directly~\cite{egele2014blanket}. 
Unfortunately, using such under-constrained execution traces directly to compute function similarities often result in many false positives~\cite{ding2019asm2vec}. For example, providing random inputs to two different functions with strict input checks might always trigger similar shallow exception handling codes and might look spuriously similar.

\vspace{.1cm}\noindent\textbf{Our approach.}
This paper presents \sys (\emph{TRansfer-learning EXecution semantics}) that trains ML models to learn the approximate execution semantics from under-constrained dynamic traces. 
Unlike prior works, which use such traces to directly measure similarity, \sys pretrains the model on diverse traces to learn each instruction's execution effect in its context.
\sys then finetunes the model by \emph{transferring} the learned knowledge from pretraining to match semantically similar functions (see Figure~\ref{fig:general}).
Our extensive experiments suggest that the approximately learned knowledge of execution semantics in pretraining significantly boosts the accuracy of matching semantically similar function binaries -- \sys excels in matching functions from different architectures, optimizations, and obfuscations.

\begin{figure*}[!t]
\centering
\includegraphics[width=\linewidth]{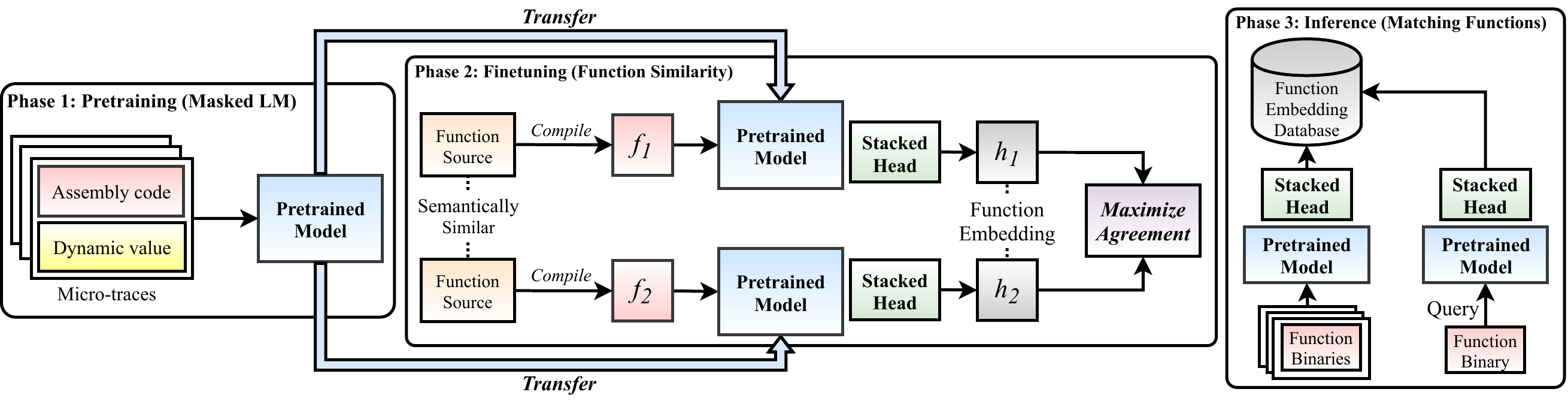}

\caption{
The workflow of \sys. We first pretrain the model on the functions' micro-traces, consisting of both instructions and dynamic values, using the masked LM task. 
We then finetune the pretrained model on the semantically similar function (only static instruction) pairs by stacking new neural network layers for function similarity tasks. 
Finetuning updates both the pretrained model and the stacked layers. 
During inference, the finetuned model computes the function embedding, whose distance encodes the function similarity.}

\label{fig:general}
\end{figure*}

Our key observation is that while under-constrained dynamic execution traces tend to contain many infeasible states, they still encode precise execution effects of many individual instructions. 
Thus, we can train an ML model to observe and learn the effect of different instructions present across a large number of under-constrained dynamic traces collected from diverse functions. 
Once the model has gained an approximate understanding of execution semantics of various instructions, we can train it to match semantically similar functions by leveraging its learned knowledge. 
As a result, during inference, we do not need to execute any functions on-the-fly while matching them~\cite{kapravelos2013revolver}, which saves significant runtime overhead.
Moreover, our trained model does not need the under-constrained dynamic traces to match functions, it only uses the function instructions, but they are \emph{augmented} with rich knowledge of execution semantics.

In this paper, we extend micro-execution~\cite{godefroid2014micro}, a form of under-constrained dynamic execution, to generate \emph{micro-traces} of a function across multiple instruction set architectures. 
A micro-trace consists of a sequence of aligned instructions and their corresponding program state values.
We pretrain the model on a large number of micro-traces gathered from diverse functions as part of training data using the masked language modeling (masked LM) task.
Notably, masked LM masks random parts in the sequence and asks the model to predict masked parts based on their context. 
This design forces the model to learn approximately how a function executes to correctly infer the missing values, which automates learning execution semantics without manual feature engineering.
Masked LM is also fully \emph{self-supervised}~\cite{devlin2018bert} -- \sys can thus be trained and further improved with arbitrary functions found in the wild.

To this end, we design a hierarchical Transformer~\cite{vaswani2017attention} that supports learning approximate execution semantics.
Specifically, our architecture models micro-trace values explicitly. 
By contrast, existing approaches often treat the numerical values as a dummy token~\cite{massarelli2019safe, ding2019asm2vec} to avoid prohibitively large vocabulary size, which cannot effectively learn the rich dependencies between concrete values that likely encode key function semantics.
Moreover, our architecture's self-attention layer is designed to model long-range dependencies in a sequence~\cite{vaswani2017attention} efficiently.
Therefore, \sys can support roughly 170$\times$ longer sequence and runs 8$\times$ faster than existing neural architectures, essential to learning embeddings of long function execution traces.

We evaluate \sys on 1,472,066 functions collected from 13 popular open-source software projects across 4 architectures (x86, x64, ARM, and MIPS) and compiled with 4 optimizations (\texttt{O0}-\texttt{O3}), and 5 obfuscation strategies~\cite{hikari}. 
\sys outperforms the state-of-the-art systems by 7.8\%, 7.2\%, and 14.3\% in matching functions across different architectures, optimizations, and obfuscations, respectively. 
Our ablation studies show that the pretraining task significantly improves the accuracy of matching semantically similar functions (by 15.7\%).
We also apply \sys in searching vulnerable functions in 180 real-world firmware images developed by well-known vendors and deployed in diverse embedded systems, including WLAN routers, smart cameras, and solar panels.
Our case study shows that \sys helps find 16 CVEs in these firmware images, which have not been disclosed in previous studies.
We make the following contributions.

\begin{itemize}
    \item We propose a new approach to matching semantically similar functions: we first train the model to learn approximate program execution semantics from micro-traces, a form of under-constrained dynamic traces, and then transfer the learned knowledge to identify semantically similar functions.
    
    
    \item We extend micro-execution to support different architectures to collect micro-traces for training. We then develop a novel neural architecture -- hierarchical Transformer -- to learn approximate execution semantics from micro-traces.

    \item We implement \sys and evaluate it on 1,472,066 functions from 13 popular software projects and libraries. \sys outperforms the state-of-the-art tools by 7.8\%, 7\%, and 14.3\%, in cross-architecture, optimization, and obfuscation function matching, respectively, while running up to 8$\times$ faster.
    Moreover, \sys helps uncover 16 vulnerabilities in 180 real-world firmware images with the latest version that are not disclosed by previous studies. 
    We release the code and dataset of \sys at \url{https://github.com/CUMLSec/trex}.

\end{itemize}

\section{Overview}
\label{sec:overview}

In this section, we use the real-world functions as motivating examples to describe the challenges of matching semantically similar functions.
We then overview our approach, focusing on how our pretraining task (masked LM) addresses the challenges.



\subsection{Challenging Cases}

\begin{figure*}[!t]
\centering

\subfloat[Cross-architecture]{
\includegraphics[width=0.3\linewidth]{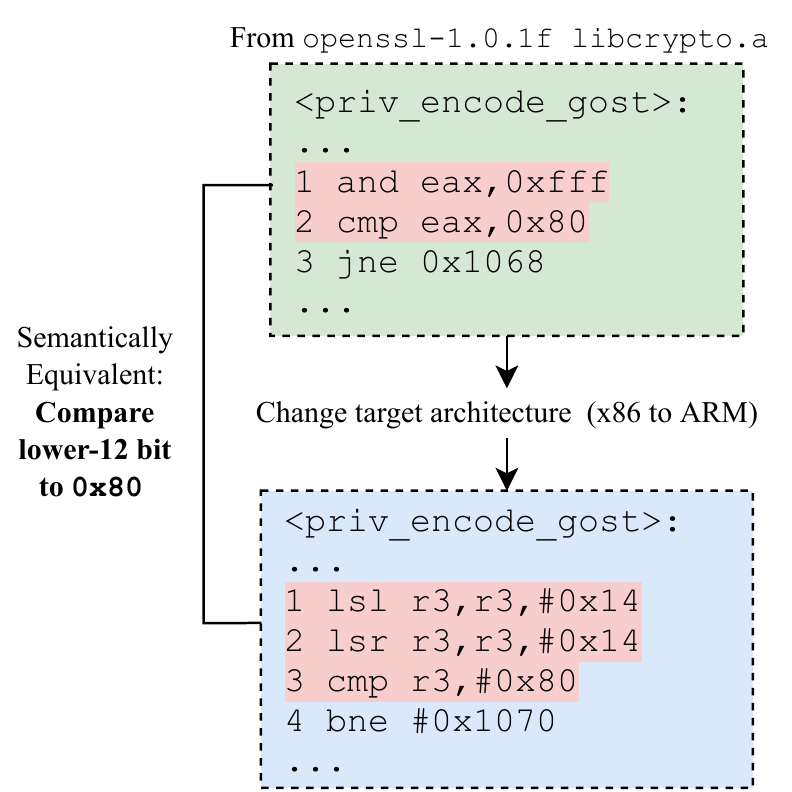}
\label{subfig:x86_arm}}
\subfloat[Cross-optimization]{
\includegraphics[width=0.33\linewidth]{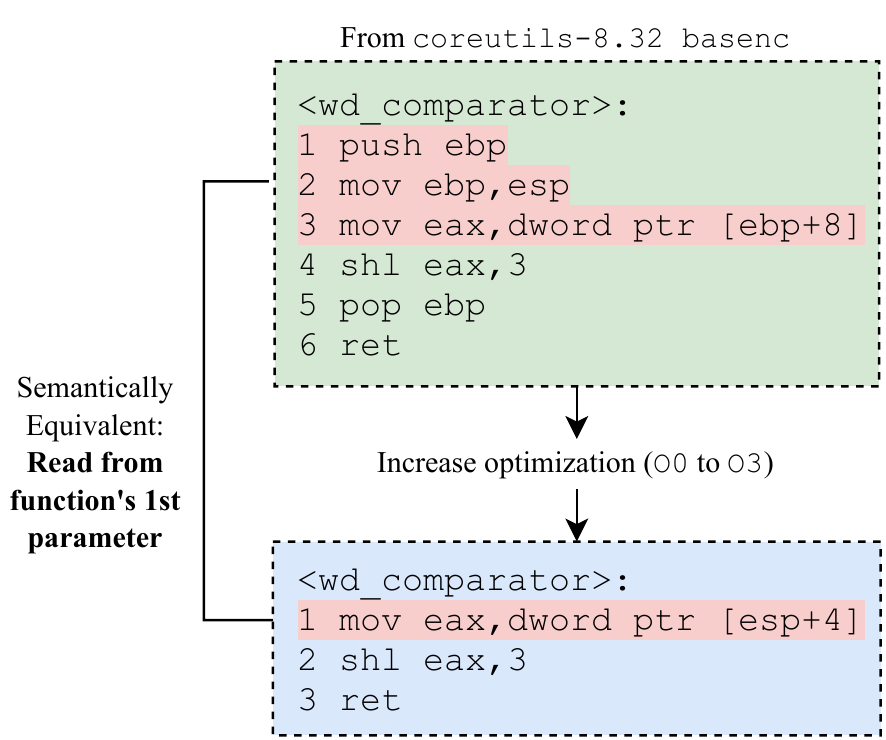}
\label{subfig:O0_O3}}
\subfloat[Cross-obfuscation]{
\includegraphics[width=0.33\linewidth]{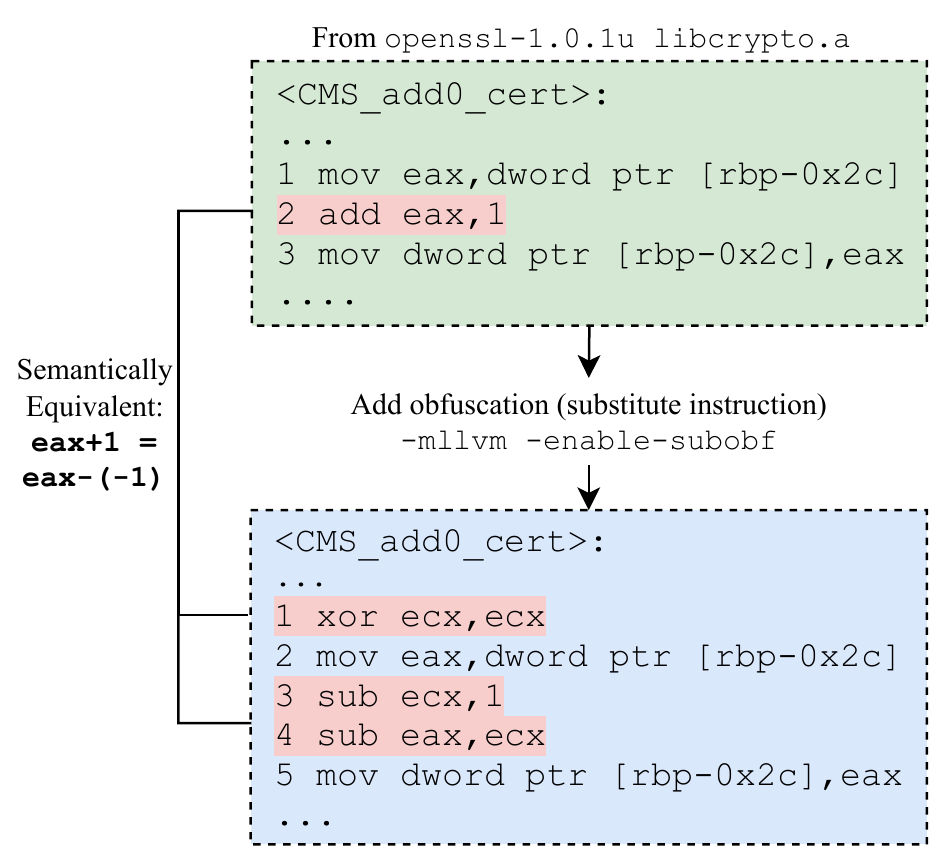}
\label{subfig:subobf}}

\caption{Challenging cases of matching semantically similar functions across different instruction architectures, optimizations, and obfuscations. (\textbf{Left}) the function \texttt{priv\_encode\_gost} is from \texttt{libcrypto.a} in \texttt{openssl-1.0.1f}. The upper function is compiled to x86 while the lower is compiled to ARM.
(\textbf{Middle}) the function \texttt{<wd\_comparator>} is from \texttt{basenc} in \texttt{coreutils-8.32}. The upper and lower function is compiled by GCC-7.5 with \texttt{-O0} and \texttt{-O3}, respectively. 
(\textbf{Right}) the function \texttt{<CMS\_add0\_cert>} is from \texttt{libcrypto.a} in \texttt{openssl-1.0.1u}. The upper function is compiled using \texttt{clang} with default options. The lower function is compiled by turning on the instruction substitution using Hikari~\cite{hikari}, \eg \texttt{-mllvm -enable-subobf}.}

\label{fig:challenge}
\end{figure*}

We use three semantically equivalent but syntactically different function pairs to demonstrate some challenges of learning from only static code.
Figure~\ref{fig:challenge} shows the (partial) assembly code of each function.

\vspace{.1cm}\noindent\textbf{Cross-architecture example.}
Consider the functions in Figure~\ref{subfig:x86_arm}. Two functions have the same execution semantics as both functions take the lower 12-bit of a register and compare it to \texttt{0x80}. 
Detecting this similarity requires understanding the approximate execution semantics of \texttt{and} in x86 and \texttt{lsl}/\texttt{lsr} in ARM. 
Moreover, it also requires understanding how the values (\ie \texttt{0xfff} and \texttt{0x14}) in the code are manipulated.
However, all existing ML-based approaches~\cite{massarelli2019safe} only learn on static code without observing each instruction's real execution effect. 
Furthermore, to mitigate the potentially prohibitive vocabulary size (\ie all possible memory addresses), existing approaches replace all register values and memory addresses with an abstract dummy symbol~\cite{massarelli2019safe,duandeepbindiff}. They thus cannot access the specific byte values to determine inherent similarity.

\vspace{.1cm}\noindent\textbf{Cross-optimization example.} 
Now consider two functions in Figure~\ref{subfig:O0_O3}. 
They are semantically equivalent as \texttt{[ebp+8]} and \texttt{[esp+4]} access the same memory location, \ie the function's first argument pushed on the stack by the caller. 
To detect such similarity, the model should understand \texttt{push} decreases the stack pointer \texttt{esp} by 4. 
The model should also notice that \texttt{mov} at line 2 assigns the decremented \texttt{esp} to \texttt{ebp} such that \texttt{ebp+8} in the upper function equals \texttt{esp+4} in the lower function.
However, such dynamic information is not reflected in the static code.

\vspace{.1cm}\noindent\textbf{Cross-obfuscation example.} 
Figure~\ref{subfig:subobf} demonstrates a simple obfuscation by instruction substitution, which essentially replaces \texttt{eax+1} with \texttt{eax-(-1)}. 
Detecting the equivalence requires understanding approximately how arithmetic operations such as \texttt{xor}, \texttt{sub}, and \texttt{add}, executes.
However, static information is not enough to expose such knowledge.


\subsection{Pretraining Masked LM on Micro-traces}
\label{subsec:overview_pretrain}

This section describes how the pretraining task, masked LM, on functions' micro-traces encourages the model to learn execution semantics.
Although it remains an open research question to explicitly prove certain knowledge is encoded by such language modeling task~\cite{saunshi2019theoretical}, we focus on describing the intuition behind the masked LM -- why predicting masked codes and values in micro-traces can help address the challenging cases in Figure~\ref{fig:challenge}. 

\vspace{.1cm}\noindent\textbf{Masked LM.} Recall the operation of masked LM: given a function's micro-trace (\ie values and instructions), we mask some random parts and train the model to predict the masked parts using those not masked. 

Note that pretraining with masked LM does not need any manual labeling effort, as it only predicts the masked part in the input micro-traces without any additional labeling effort.
Therefore, \sys can be trained and further improved with a substantial number of functions found in the wild.
The benefit of this is that a certain instruction not micro-executed in one function is highly likely to appear in at least one of the other functions’ micro-traces, supporting \sys to approximate diverse instructions' execution semantics.



\vspace{.1cm}\noindent\textbf{Masking register.} Consider the functions in Figure~\ref{subfig:subobf}, where they essentially increment the value at stack location \texttt{[rbp-0x2c]} by 1. 
The upper function directly loads the value to \texttt{eax}, increments by 1, and stores the value in \texttt{eax} back to stack.
The lower function, by contrast, takes a convoluted way by first letting \texttt{ecx} to hold the value -1, and decrements \texttt{eax} by \texttt{ecx}, and stores the value in \texttt{eax} back to stack.

We mask the \texttt{eax} at line 3 in the upper function. We find that our pretrained model can correctly predict its name and dynamic value.
This implies the model understands the semantics of \texttt{add} and can deduce the value of \texttt{eax} in line 3 after observing the value of \texttt{eax} in line 2 (before the addition takes the effect).
We also find the model can recover the values of masked \texttt{ecx} in line 4 and \texttt{eax} in line 5, implying the model understands the execution effect of \texttt{xor} and \texttt{sub}.

The understanding of such semantics can significantly improve the robustness in matching similar functions -- when finetuned to match similar functions, the model is more likely to learn to attribute the similarity to their \emph{similar execution effects}, instead of their syntactic similarity.

\vspace{.1cm}\noindent\textbf{Masking opcode.}
Besides masking the register and its value, we can also mask the opcode of an instruction. Predicting the opcode requires the model to understand the execution effect of each opcode.
Consider Figure~\ref{subfig:O0_O3}, where we mask \texttt{mov} in line 2 of upper function.
We find our pretrained model predicts \texttt{mov} with the largest probability (larger than the other potential candidates such as \texttt{add}, \texttt{inc}, etc.).

To correctly predict the opcode, the model should have learned several key aspects of the function semantics. 
First, according to its context, \ie the value of \texttt{ebp} at line 3 and \texttt{esp} at line 2, it learns \texttt{mov} is most probable as it assigns the value of \texttt{esp} to \texttt{ebp}. Other opcodes are less likely as their execution effect conflicts with the observed resulting register values.
This also implicitly implies the model learns the approximate execution semantics of \texttt{mov}.
Second, the model also learns the common calling convention and basic syntax of x86 instructions, \eg only a subset of opcodes accept two operands (\texttt{ebp,esp}). It can thus exclude many syntactically impossible opcodes such as \texttt{push}, \texttt{jmp}, etc.

The model can thus infer \texttt{ebp} (line 3 of upper function) equals to \texttt{esp}. The model may have also learned \texttt{push} decrements stack pointer \texttt{esp} by 4 bytes, from other masked samples.
Therefore, when the pretrained model is finetuned to match the two functions, the model is more likely to learn that the semantic equivalence is due to that \texttt{[ebp+8]} in the upper function and \texttt{[esp+4]} in the lower function refer to the same address, instead of their similar syntax.

\vspace{.1cm}\noindent\textbf{Other masking strategies.}
Note that we are not constrained by the number or the type of items (\ie register, opcode, etc.) in the instructions to mask, \ie we can mask complete instructions or even a consecutive sequence of instructions, and we can mask dynamic values of random instructions’ input-output. 
Moreover, the masking operation dynamically selects random subsets of code blocks and program states at each training iteration and on different training samples.
As a result, it enables the model to learn the diverse and composite effect of the instruction sequence, essential to detecting similarity between functions with various instructions.
In this paper, we adopt a completely randomized strategy to choose what part of the micro-trace to mask with a fixed masking percentage (see Section~\ref{subsec:pretrain_method} for details).
However, we envision a quite interesting future work to study a better (but still cheap) strategy to dynamically choose where and how much to mask.

\section{Threat Model}
\label{sec:threat}

We assume no access to the debug symbols or source while comparing binaries.
Indeed, there exist many approaches to reconstruct functions from stripped binaries~\cite{bao2014byteweight, shin2015recognizing, andriesse2017compiler, di2017rev, pei2021xda}. 
Moreover, we assume the binary can be readily disassembled, \ie it is not packed nor transformed by virtualization-based obfuscator~\cite{vmprotect, ugarte2015sok}.

\vspace{.1cm}\noindent\textbf{Semantic similarity.}
We consider two semantically similar functions as having the same input-output behavior (\ie given the same input, two functions produce the same output).
Similar to previous works~\cite{xu2017neural, massarelli2019safe, ding2019asm2vec}, we treat functions compiled from the same source as similar, regardless of architectures, compilers, optimizations, and obfuscation transforms.

\section{Methodology}
\label{sec:method}

This section describes \sys's design specifics, including our micro-tracing semantics, our learning architecture's details, and pretraining and finetuning workflow.

\subsection{Micro-tracing Semantics}
\label{subsec:microx-semantics}

We implement micro-execution by Godefroid~\cite{godefroid2014micro} to handle x64, ARM, and MIPS, where the original paper only describes x86 as the use case.
In the following, we briefly explain how we micro-execute an individual function binary, highlighting the key algorithms in handling different types of instructions.

\vspace{.1cm}\noindent\textbf{IR Language.}
To abstract away the complexity of different architectures' assembly syntax, we introduce a low-level intermediate representation (IR) that models function assembly code. 
We only include a subset of the language specifics to illustrate the implementation algorithm.
Figure~\ref{fig:microx-syntax} shows the grammar of the IR.
Note that the IR here only serves to facilitate the discussion of our micro-tracing implementation. 
In our implementation, we use real assembly instructions and tokenize them as model’s input (Section~\ref{subsec:input_repr}).

Notably, we denote memory reads and writes by \texttt{load($e$)} and \texttt{store($e_{v},e_{a}$)} (\ie store the value expression $e_v$ to address expression $e_a$), which generalize from both the load-store architecture (\ie ARM, MIPS) and register-memory architecture (\ie x86). 
Both operations can take as input $e$ -- an expression that can be an explicit hexadecimal number (denoting the address or a constant), a register, or a result of an operation on two registers.
We use \texttt{jmp} to denote the general jump instruction, which can be both direct or indirect jump (\ie the expression $e_a$ can be a constant $c$ or a register $r$).
The jump instruction can also be unconditional or conditional. Therefore, the first parameter in \texttt{jmp} is the conditional expression $e_c$ and unconditional jump will set $e_c$ to \texttt{true}.
We represent function invocations and returns by \texttt{call} and \texttt{ret}, where \texttt{call} is parameterized by an expression, which can be an address (direct call) or a register (indirect call).

\begin{figure}[!t]
\centering

\includegraphics[width=\linewidth]{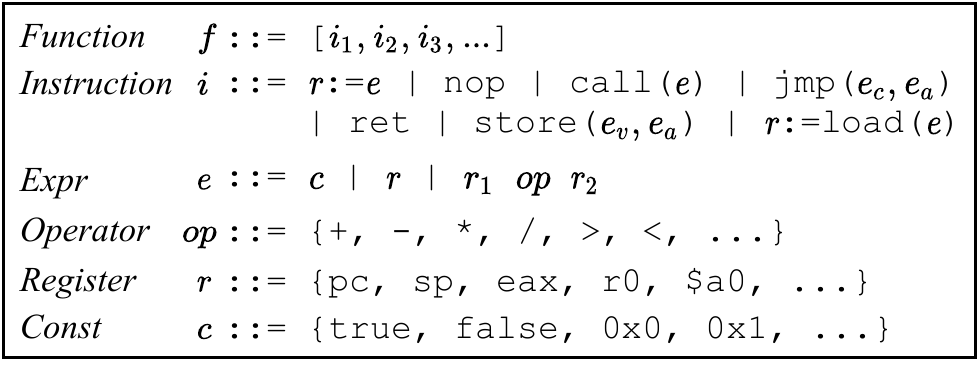}

\caption{Low-level IR for representing assembly code. The IR abstracts away the actual assembly syntax that are disparate across different architectures.}
\label{fig:microx-syntax}
\end{figure}

\vspace{.1cm}\noindent\textbf{Micro-tracing algorithm.} 
Algorithm~\ref{alg:microx} outlines the basic steps of micro-tracing a given function $f$.
First, it initializes the memory to load the code and the corresponding stack.
It then initializes all registers except the special-purpose register, such as the stack pointer or the program counter.
Then it starts linearly executing instructions of $f$. 
We map the memory address \emph{on-demand} if the instruction access the memory (\ie read/write).
If the instruction reads from memory, we further initialize a random value in the specific memory addresses.
For call/jump instructions, we first examine the target address and skip the invalid jump/call, known as ``forced execution''~\cite{peng2014x}.
By skipping unreachable jumps and calls, it can keep executing the function till the end of the function and exposes more behaviors, \eg skipping potential input check exceptions.
Since the \texttt{nop} instructions can serve as padding between instructions within a function, we simply skip \texttt{nop}.
We terminate the micro-tracing when it finishes executing all instructions, reaches \texttt{ret}, or times out.
Figure~\ref{fig:arithmetic} and~\ref{fig:stack} demonstrate sample micro-traces of real-world functions.




\subsection{Input Representation}
\label{subsec:input_repr}

Formally, given a function $f$ (\ie assembly code) and its micro-trace $t$ (by micro-executing $f$), we prepare the model input $x$, consisting of 5 types of token sequence with the same size $n$. 
Figure~\ref{fig:arch} shows the model input example and how they are masked and processed by the hierarchical Transformer to predict the corresponding output as a pretraining task. 

\begin{algorithm}[!t]
\footnotesize
\setlength{\tabcolsep}{2pt}
	\caption{Micro-tracing a function $f$}
\label{alg:microx}
\begin{tabular}{lp{2.9in}}
\textbf{Input}:
    & Function binary $f$. All registers $r$. \\
\textbf{Output}:
    & Micro-trace $t$.
\end{tabular}

\begin{spacing}{1.1}
\begin{algorithmic}[1]

\State $\mathbb{I}\gets get\_instructions(f)$ \Comment{put all instructions in $f$ into a queue}
\State $t\gets empty\_vector$
\State \texttt{sp} $\gets init\_stack\_pointer\_addr()$ \Comment{stack pointer address}
\State \texttt{pc} $\gets init\_program\_counter\_addr()$ \Comment{first instruction's address}
\State \texttt{sm} $\gets mem\_map(\texttt{sp}, \texttt{STACK\_SIZE})$ \Comment{initialize stack memory}
\State \texttt{cm} $\gets mem\_map(\texttt{pc}, |\mathbb{I}|)$ \Comment{initialize memory for code}
\vspace{.1cm}
\For{each register $r_i$ in $r$\textbackslash\{\texttt{sp},\texttt{pc}\}}
    \State $r_i\gets $ \texttt{random\_init()} \Comment{initialize register values}
\EndFor

\While{$\mathbb{I}\neq \emptyset$}
    \State $i$ $\leftarrow$ $dequeue(\mathbb{I})$
	\If{$i.type$ = \texttt{load} \textbf{or} $i.type$ = \texttt{store}} \Comment{memory access}
	    \State $mem\_map(i.access\_addr, i.access\_size)$
	    \If{$i.type$ = \texttt{load}}
	        \State $write\_random(i.access\_addr)$ 
	    \EndIf
	    \State $t\gets t\cup execute(i)$
	\ElsIf{$i.type$ = \texttt{jmp} \textbf{or} $i.type$ = \texttt{call}} \Comment{control transfer}
	    \If{$i.target\_addr \not\in [cm.min\_addr,cm.max\_addr]$}
	        \State \textbf{continue} \Comment{skip illegal jump/call}
	    \EndIf
	    \State $t\gets t\cup execute(i)$
	\ElsIf{$i.type$ = \texttt{nop}} \Comment{NOP}
	    \State \textbf{continue}
	\ElsIf{$i.type$ = \texttt{ret}} \Comment{return}
	    \State \textbf{break}
	\Else \Comment{all other instructions}
	    \State $t\gets t\cup execute(i)$
	\EndIf
\EndWhile


\end{algorithmic}
\end{spacing}
\end{algorithm}

\vspace{.1cm}\noindent\textbf{Micro-trace code sequence.}
The first sequence $x_f$ is the assembly code sequence: $x_f=\{mov, eax, +, ...\}^n$, generated by tokenizing the assembly instructions in the micro-trace. 
We treat all symbols appear in the assembly instructions as tokens. 
Such a tokenization aims to preserve the critical hint of the syntax and semantics of the assembly instructions.
For example, we consider even punctuation to be one of the tokens, \eg ``,'', ``['', ``]'', as ``,'' implies the token before and after it as destination and source of \texttt{mov} (in Intel syntax), respectively, and ``['' and ``]'' denote taking the address of the operands reside in between them.

We take special treatment of numerical values appear in the assembly code.
Treating numerical values as regular text tokens can incur prohibitively large vocabulary size, \eg $2^{32}$ number of possibilities on 32-bit architectures.
To avoid this problem, we move all numeric values to the micro-trace value sequence (that will be learned by an additional neural network as detailed in the following) and replace them with a special token \texttt{num} (\eg last token of input in Figure~\ref{fig:arch}). 
With all these preprocessing steps, the vocabulary size of $x_f$ across all architectures is 3,300.

\vspace{.1cm}\noindent\textbf{Micro-trace value sequence.}
The second sequence $x_t$ is the micro-trace value sequence, where each token in $x_t$ is the dynamic value from micro-tracing the corresponding code.
As discussed in Section~\ref{sec:overview}, we keep \emph{explicit} values (instead of a dummy value used by existing approaches) in $x_t$.
Notably, we use the dynamic value for each token (\eg register) in an instruction before it is executed.
For example, in \texttt{mov eax,0x8; mov eax,0x3}, the dynamic value of the second \texttt{eax} is \texttt{0x8}, as we take the value of \texttt{eax} before \texttt{mov eax,0x3} is executed.
For code token without dynamic value, \eg \texttt{mov}, we use dummy values (see below). 

\vspace{.1cm}\noindent\textbf{Position sequences.}
The position of each code and value token is critical for inferring binary semantics. Unlike natural language, where swapping two words can roughly preserve the same semantic meaning, swapping two operands can significantly change the instructions.
To encode the inductive bias of position into our model, we introduce \emph{instruction position sequence} $x_c$ and \emph{opcode/operand position sequence} $x_o$ to represent the relative positions between the instructions and within each instruction.
As shown in Figure~\ref{fig:arch}, $x_c$ is a sequence of integers encoding the position of each instruction. All opcodes/operands within a single instruction share the same value.
$x_o$ is a sequence of integers encoding the position of each opcode and operands within a single instruction.

\vspace{.1cm}\noindent\textbf{Architecture sequence.}
Finally, we feed the model with an extra sequence $x_a$, describing the input binary's instruction set architecture. The vocabulary of $x_a$ consists of 4 architectures: $x_a=\{$x86, x64, ARM, MIPS$\}^n$.
This setting helps the model to distinguish between the syntax of different architecture.

\begin{figure}[!t]
\centering

\includegraphics[width=\linewidth]{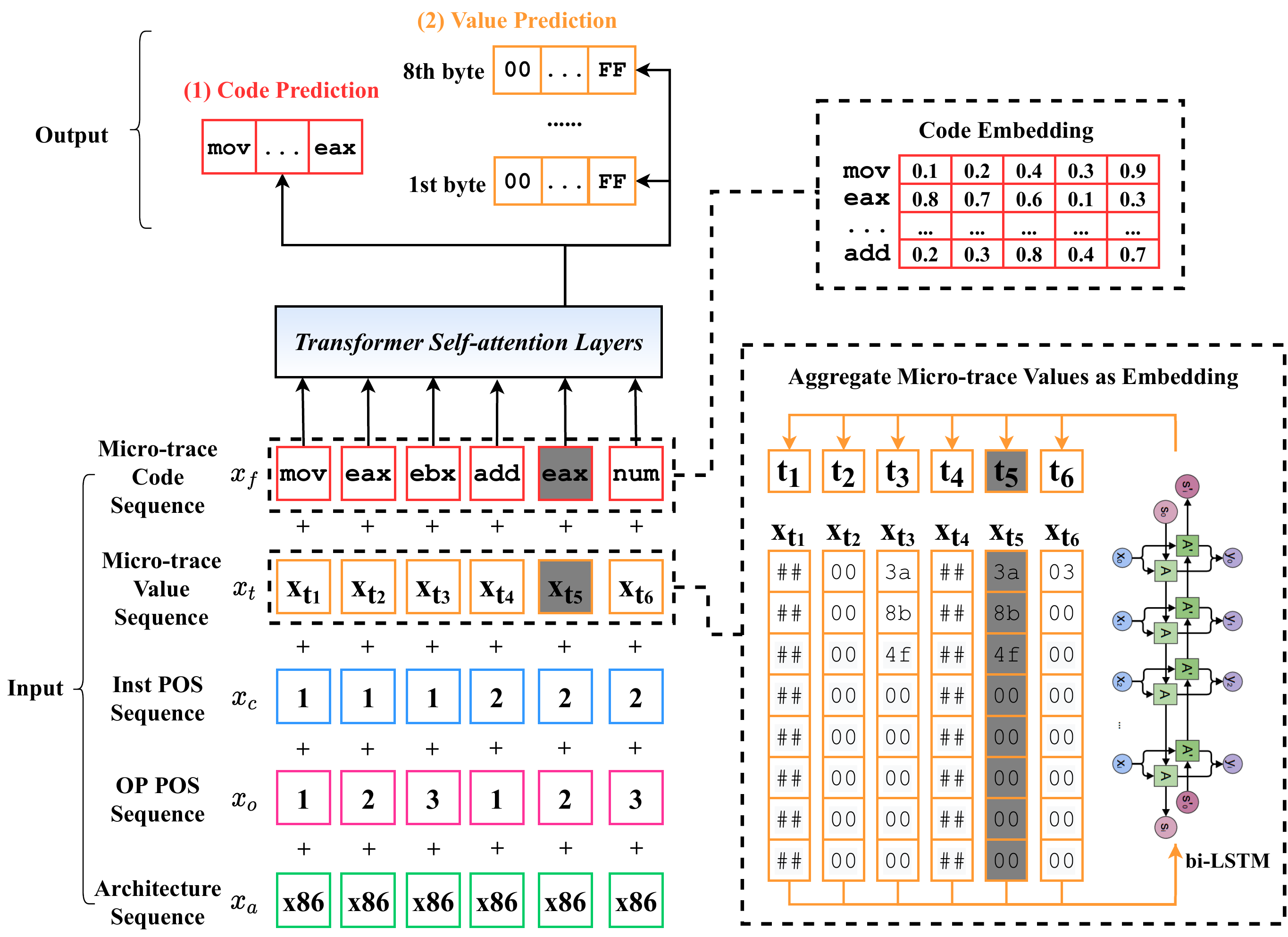}

\caption{Model architecture with input-output example. The masked input is marked in gray. In pretraining, the loss function consists of the cross-entropy losses of both (1) code prediction and (2) value prediction, where the value prediction consists of predicting each of the 8 bytes of the micro-trace value.}
\label{fig:arch}
\end{figure}

\vspace{.1cm}\noindent\textbf{Encoding numeric values.}
As mentioned above, treating concrete values as independent tokens can lead to prohibitively large vocabulary size. 
We design a \emph{hierarchical input encoding scheme} to address this challenge.
Specifically, let $x_{t_i}$ denote the $i$-th value in $x_t$.
We represent $x_{t_i}$ as an (padded) 8-byte fixed-length byte sequence $x_{t_i}=$\{\texttt{0x00}, ..., \texttt{0xff}\}$^8$ ordered in Big-Endian.
We then feed $x_{t_i}$ to a 2-layer bidirectional LSTM (bi-LSTM) and take its last hidden cell's embedding as the value representation $t_i=bi$-$LSTM(x_{t_i})$.
Here $t_i$ denotes the output of applying the embedding to $x_{t_i}$.
To make the micro-trace code tokens without dynamic values (\eg opcode) align with the byte sequence, we use a dummy sequence (\texttt{\#\#}) with the same length.
Figure~\ref{fig:arch} (right-hand side) illustrates how bi-LSTM takes the byte sequence and computes the embedding.

Such a design significantly reduces the vocabulary size as now we have only 256 possible byte values to encode.
Moreover, bi-LSTM encodes the potential dependencies between high and low bytes within a single value.
This setting thus supports learning better relationships between different dynamic values (\eg memory region and offset) as opposed to treating all values as dummy tokens~\cite{shi2019learning}. 

\subsection{Pretraining with Micro-traces}
\label{subsec:pretrain_method}

This section describes the pretraining task on the function binaries and their micro-traces, focusing on how the model masks its input $x$ (5 sequences) and predicts the masked part to learn the approximate execution semantics.

\vspace{.1cm}\noindent\textbf{Learned embeddings.}
We embed each token in the 5 sequences with the same embedding dimension $d_{emb}$ so that they can be summed as a single sequence as input to the Transformer.
Specifically, let $E_f(x_f)$, $E_t(x_t)$, $E_c(x_c)$, $E_o(x_o)$, $E_a(x_a)$ denote applying the embedding to the tokens in each sequence, respectively. 
We have $E_{i}$, the embedding of $x_i$:
\begin{equation*}
    E_{i} = E_f(x_{f_i})+E_t(x_{t_i})+E_c(x_{c_i})+E_o(x_{o_i})+E_a(x_{a_i})
\label{eq:pos}
\end{equation*}

Here $x_{f_i}$ denote the $i$-th token in $x_f$, where other sequence (\ie $x_t$, $x_c$, $x_o$, $x_a$) follow the similar denotation.
Note that $E_t(x_{t_i})$ leverages the bi-LSTM to encode the byte sequences (see Section~\ref{subsec:input_repr}), while the others simply multiplies the token (\ie one-hot encoded~\cite{harris2010digital}) with an embedding matrix.
Figure~\ref{fig:arch} (right-hand side) illustrates the two embedding strategies.

\vspace{.1cm}\noindent\textbf{Masked LM.}
We pretrain the model using the masked LM objective.
Formally, for each sequence $(x_1,x_2,...,x_n)$ in a given training set, we randomly mask out a selected percentage of tokens in each sequence. 
Specifically, we mask the code token and value token in $x_f$ and $x_t$, respectively, and replace them with a special token \texttt{<MASK>} (marked gray in Figure~\ref{fig:arch}).
As masked LM trains on micro-traces without requiring additional labels, it is fully unsupervised.

Let $m(E_i)$ denote the embedding of the masked $x_i$ and $\mathbb{MP}$ a set of positions on which the masks are applied. 
The model $g_p$ (to be pretrained) takes as input a sequence of embeddings with random tokens masked: $(E_1,..., m(E_i),...,E_n), i\in \mathbb{MP}$, and predicts the code and the values of the masked tokens: $\{\hat{x}_{f_i}, \hat{v}_i|i\in \mathbb{MP}\}=g_p(E_1,..., m(E_i),...,E_n)$.
Let $g_p$ be parameterized by $\theta$, the objective of training $g_p$ is thus to search for $\theta$ that minimizes the cross-entropy losses between (1) the predicted masked code tokens and the actual code tokens, and (2) predicted masked values (8 bytes) and the actual values. For ease of exposition, we omit summation over all samples in the training set.

\begin{equation}
\label{eq:pretrain}
\centering
    \argmin_{\theta} \sum\limits_{i=1}^{|\mathbb{MP}|} (-x_{f_i}\log(\hat{x}_{f_i})+ \alpha\sum\limits_{j=1}^{8}-x_{t_{ij}}\log(\hat{x}_{t_{ij}}))
\end{equation}

$\hat{x}_{t_{ij}}$ denotes the predicted $j$-th byte of $x_{t_i}$ (the $i$-th token in $x_t$).
$\alpha$ is a hyperparameter that weighs the cross-entropy losses between predicting code tokens and predicting values.

\vspace{.1cm}\noindent\textbf{Masking strategy.}
For each chosen token to mask, we randomly choose from the masking window size from $\{1,3,5\}$, which determines how many consecutive neighboring tokens of the chosen tokens are also masked~\cite{joshi2020spanbert}.
For example, if we select $x_5$ (the 5-th token) in Figure~\ref{fig:arch} to mask and the masking window size is 3, $x_4$ and $x_6$ will be masked too.
We then adjust accordingly to ensure the overall masked tokens still account for the initially-chosen masking percentage.

\vspace{.1cm}\noindent\textbf{Contextualized embeddings.}
We employ the self-attention layers~\cite{vaswani2017attention} to endow contextual information to each embedding $E_i$ of the input token.
Notably, let $E_{l}=(E_{l,1},...,E_{l,n})$ denote the embeddings produced by the $l$-th self-attention layer.
We denote the embeddings before the model ($E_i$ as defined in Section~\ref{subsec:input_repr}) as $E_{0,i}$.
Each token's embedding at each layer will attend to all other embeddings, aggregate them, and update its embedding in the next layer.
The embeddings after each self-attention layer are known as \emph{contextualized embeddings}, which encodes the context-sensitive meaning of each token (\eg \texttt{eax} in \texttt{mov eax,ebx} has different embedding with that in \texttt{jmp eax}).
This is in contrast with static embeddings, \eg word2vec commonly used in previous works~\cite{duandeepbindiff, ding2019asm2vec}, where a code token is assigned to a fixed embedding regardless of the changed context.

The learned embeddings $E_{l,i}$ after the last self-attention layer encodes the approximate execution semantics of each instruction and the overall function.
In pretraining, $E_{l,i}$ is used to predict the masked code. While in finetuning, it will be leveraged to match similar functions (Section~\ref{subsec:finetuning_method}).


\subsection{Finetuning for Function Similarity}
\label{subsec:finetuning_method}

Given a function pair, we feed each function's \emph{static code} (instead of micro-trace as discussed in Section~\ref{sec:intro}) to the pretrained model $g_p$ and obtain the pair of embedding sequences produced by the last self-attention layer of $g_p$: $E^{(1)}_{k}=(E^{(1)}_{k,1},...,E^{(1)}_{k,n})$ and $E^{(2)}_{k}=(E^{(2)}_{k,1},...,E^{(2)}_{k,n})$ where $E^{(1)}_{k}$ corresponds to the first function and $E^{(2)}_{k}$ corresponds to the second.
Let $y=\{-1,1\}$ be the ground-truth indicating the similarity (1 -- similar, -1 -- dissimilar) between two functions.
We stack a 2-layer Multi-layer Perceptrons $g_t$, taking as input the average of embeddings for each function, and producing a function embedding: 
\begin{equation*}
g_t(E_{k})=tanh((\sum\limits_{i=1}^n E_{k,i})/n)\cdot W_1)\cdot W_2
\end{equation*}

Here $W_1\in \mathbb{R}^{d_{emb}\times d_{emb}}$ and $W_2\in \mathbb{R}^{d_{emb}\times d_{func}}$ transforms the average of last self-attention layers embeddings $E_k$ with dimension $d_{emb}$ into the function embedding with the dimension $d_{func}$. 
$d_{func}$ is often chosen smaller than $d_{emb}$ to support efficient large-scale function searching~\cite{xu2017neural}.
Now let $g_t$ be parameterized by $\theta$, the finetuning objective is to minimize the cosine embedding loss ($l_{ce}$) between the ground-truth and the cosine distance between two function embeddings:
\begin{equation*}
  \argmin_{\theta}\ l_{ce}(g_t(E^{(1)}_{k}),g_t(E^{(2)}_{k}), y)
\end{equation*}  
where
\begin{equation}
\label{eq:cosembloss}
l_{ce}(x_1,x_2,y) = \left\{\begin{matrix}
  1-cos(x_1,x_2) & y=1\\ 
  max(0,cos(x_1,x_2)-\xi) & y=-1
  \end{matrix}\right.
\end{equation}
$\xi$ is the margin usually chosen between 0 and 0.5~\cite{paszke2019pytorch}.
As both $g_p$ and $g_t$ are neural nets, optimizing Equation~\ref{eq:pretrain} and Equation~\ref{eq:cosembloss} can be guided by gradient descent via backpropagation.

After finetuning $g_t$, the 2-layer multilayer perceptrons, and $g_p$, the pre-trained model, we compute the function embedding $f_{emb}=g_t(g_p(f))$ and the similarity between two functions is measured by the cosine similarity between two function embedding vectors: $cos(f^{(1)}_{emb},f^{(2)}_{emb})$.

\section{Implementation and Experimental Setup}
\label{sec:impl}

We implement \sys using fairseq, a sequence modeling toolkit~\cite{ott2019fairseq}, based on PyTorch 1.6.0 with CUDA 10.2 and CUDNN 7.6.5.
We run all experiments on a Linux server running Ubuntu 18.04, with an Intel Xeon 6230 at 2.10GHz with 80 virtual cores including hyperthreading, 385GB RAM, and 8 Nvidia RTX 2080-Ti GPUs. 

\vspace{.1cm}\noindent\textbf{Datasets.}
To train and evaluate \sys, we collect 13 popular open-source software projects. 
These projects include Binutils-2.34, Coreutils-8.32, Curl-7.71.1, Diffutils-3.7, Findutils-4.7.0, GMP-6.2.0, ImageMagick-7.0.10, Libmicrohttpd-0.9.71, LibTomCrypt-1.18.2, OpenSSL-1.0.1f and OpenSSL-1.0.1u, PuTTy-0.74, SQLite-3.34.0, and Zlib-1.2.11. 
We compile these projects into 4 architectures \ie x86, x64, ARM (32-bit), and MIPS (32-bit), with 4 optimization levels (OPT), \ie \texttt{O0}, \texttt{O1}, \texttt{O2}, and \texttt{O3}, using \texttt{GCC-7.5}.
Specifically, we compile the software projects based on its makefile, by specifying \texttt{CFLAGS} (to set optimization flag), \texttt{CC} (to set cross-compiler), and \texttt{--host} (to set the cross-compilation target architecture).
We always compile to dynamic shared objects, but resort to static linking when we encounter build errors.
We are able to compile all projects with these treatments.

We also obfuscate all projects using 5 types of obfuscations (OBF) by Hikari~\cite{hikari} on x64 -- an obfuscator based on \texttt{clang-8}.
The obfuscations include bogus control flow (\texttt{bcf}), control flow flattening (\texttt{cff}), register-based indirect branching (\texttt{ibr}), basic block splitting (\texttt{spl}), and instruction substitution (\texttt{sub}).
As we encounter several errors in cross-compilation using Hikari (based on Clang)~\cite{hikari}, and the baseline system (\ie Asm2Vec~\cite{ding2019asm2vec}) to which we compare only evaluates on x64, we restrict the obfuscated binaries for x64 only.
As a result, we have 1,472,066 functions, as shown in Table~\ref{tab:dataset}. 

\begin{table*}[!t]

\footnotesize
\setlength{\tabcolsep}{2.2pt}
\centering
\renewcommand{\arraystretch}{1}
\setlength\aboverulesep{0.2pt}
\setlength\belowrulesep{0.2pt}

\caption{Number of functions for each project across 4 architectures with 4 optimization levels and 5 obfuscations.
}
\label{tab:dataset}

\rowcolors{1}{}{lightblue} 

\begin{tabular}{c|c|l|l|l|l|l|l|l|l|l|l|l|l|l?{1.1pt}r}
\toprule[1.1pt]
\multirow{2}{*}{ARCH}  & \multirow{2}{*}{\begin{tabular}[c]{@{}l@{}}OPT\\ OBF\end{tabular}} & \multicolumn{14}{c}{\bf \# Functions}                             \\ \cline{3-16} 
 &
   &
  \multicolumn{1}{c|}{Binutils} &
  \multicolumn{1}{c|}{Coreutils} &
  \multicolumn{1}{c|}{Curl} &
  \multicolumn{1}{c|}{Diffutils} &
  \multicolumn{1}{c|}{Findutils} &
  \multicolumn{1}{c|}{GMP} &
  \multicolumn{1}{c|}{ImageMagick} &
  \multicolumn{1}{c|}{Libmicrohttpd} &
  \multicolumn{1}{c|}{LibTomCrypt} &
  \multicolumn{1}{c|}{OpenSSL} &
  \multicolumn{1}{c|}{PuTTy} &
  \multicolumn{1}{c|}{SQLite} &
  \multicolumn{1}{c?{1.1pt}}{Zlib} &
  \multicolumn{1}{c}{\textbf{Total}} \\ \midrule[.9pt]
  & \texttt{O0}                  & 25,492   & 19,992   & 1,067   & 944   & 1,529   & 766   & 2,938   & 200   & 779   & 11,918   & 7,087   & 2,283   & 157   & 75,152    \\ \cline{2-16} 
                      & \texttt{O1}                   & 20,043   & 14,918   & 771   & 694   & 1,128   & 704   & 2,341   & 176   & 745   & 10,991   & 5,765   & 1,614   & 143   & 60,033    \\ \cline{2-16} 
                      & \texttt{O2}                   & 19,493   & 14,778   & 765   & 693   & 1,108   & 701   & 2,358   & 171   & 745   & 11,001   & 5,756   & 1,473   & 138   & 59,180    \\ \cline{2-16} 
                      & \texttt{O3}                   & 17,814   & 13,931   & 697   & 627   & 983   & 680   & 2,294   & 160   & 726   & 10,633   & 5,458   & 1,278   & 125   & 55,406    \\ \cline{2-16} 
\multirow{-5}{*}{ARM}                      & \multicolumn{14}{c?{1.1pt}}{Total \# Functions of ARM}  & \textbf{249,771}   \\ \midrule[.5pt]
\multirow{5}{*}{MIPS} & \texttt{O0}                   & 28,460   & 18,843   & 1,042   & 906   & 1,463   & 734   & 2,840   & 200   & 779   & 11,866   & 7,003   & 2,199   & 153   & 76,488    \\ \cline{2-16} 
                      & \texttt{O1}                   & 22,530   & 13,771   & 746   & 653   & 1,059   & 670   & 2,243   & 176   & 745   & 10,940   & 5,685   & 1,530   & 139   & 60,887    \\ \cline{2-16} 
                      & \texttt{O2}                   & 22,004   & 13,647   & 741   & 653   & 1,039   & 667   & 2,260   & 171   & 743   & 10,952   & 5,677   & 1,392   & 135   & 60,081   \\ \cline{2-16} 
                      & \texttt{O3}                   & 20,289   & 12,720   & 673   & 584   & 917   & 646   & 2,198   & 161   & 724   & 10,581   & 5,376   & 1,197   & 121   & 56,187   \\ \cline{2-16} 
                      & \multicolumn{14}{c?{1.1pt}}{Total \# Functions of MIPS} & \textbf{253,643}   \\ \midrule[.5pt]
  & \texttt{O0}                   & 37,783   & 24,383   & 1,335   & 1,189   & 1,884  & 809   & 3,876   & 326   & 818   & 12,552   & 7,548   & 2,923   & 204   & 95,630   \\ \cline{2-16} 
                      & \texttt{O1}                   & 32,263   & 20,079   & 1,013   & 967   & 1,516   & 741   & 3,482   & 280   & 782   & 11,578   & 6,171   & 2,248   & 196   & 81,316    \\ \cline{2-16} 
                      & \texttt{O2}                   & 32,797   & 21,082   & 1,054   & 1,006   & 1,524   & 728   & 3,560   & 265   & 784   & 11,721   & 6,171   & 2,113   & 183   & 82,988   \\ \cline{2-16} 
                      & \texttt{O3}                   & 34,055   & 22,482   & 1,020   & 1,052   & 1,445   & 707   & 3,597   & 284   & 760   & 11,771   & 5,892   & 1,930   & 197   & 85,192   \\ \cline{2-16} 
\multirow{-5}{*}{x86}                      & \multicolumn{14}{c?{1.1pt}}{Total \# Functions of x86} & \textbf{358,261}   \\ \midrule[.5pt]
\multirow{5}{*}{x64}  & \texttt{O0}                   & 26,757   & 17,238   & 1,034   & 845   & 1,386   & 751   & 2,970   & 200   & 782   & 12,047   & 7,061   & 2,190   & 151   & 73,412    \\ \cline{2-16} 
                      & \texttt{O1}                   & 21,447   & 12,532   & 739   & 600   & 1,000   & 691   & 2,358   & 176   & 745   & 11,120   & 5,728   & 1,523   & 137   & 58,796    \\ \cline{2-16} 
                      & \texttt{O2}                   & 20,992   & 12,206   & 734   & 596   & 976   & 689   & 2,374   & 171   & 742   & 11,136   & 5,703   & 1,380   & 132   & 57,831    \\ \cline{2-16} 
                      & \texttt{O3}                   & 19,491   & 11,488   & 662   & 536   & 857   & 667   & 2,308   & 160   & 725   & 10,768   & 5,390   & 1,183   & 119   & 54,354   \\ \cline{2-16} 
                      & \multicolumn{14}{c?{1.1pt}}{Total \# Functions of x64}  & \textbf{244,393}   \\ \midrule[1.1pt]
  & \texttt{bcf}                   & 27,734   & 17,093   & 998   & 840   & 1,388   & 746   & 2,833   & 200   & 782   & 10,768   & 7,069   & 2,183   & 151   & 72,785    \\ \cline{2-16} 
                      & \texttt{cff}                   & 27,734   & 17,093   & 998   & 840   & 1,388   & 746   & 2,833   & 200   & 782   & 10,903   & 7,069   & 2,183   & 151   & 72,920    \\ \cline{2-16}
                  & \texttt{ibr}                   & 27,734   & 17,105   & 998   & 842   & 1,392   & 746   & 2,833   & 204   & 782   & 12,045   & 7,069   & 2,183   & 151   & 74,084    \\ \cline{2-16}  
                      & \texttt{spl}                   & 27,734   & 17,093   & 998   & 840   & 1,388   & 746   & 2,833   & 200   & 782   & 10,772   & 7,069   & 2,183   & 151   & 72,789   \\ \cline{2-16}  
                      & \texttt{sub}                   & 27,734   & 17,093   & 998   & 840   & 1,388   & 746   & 2,833   & 200   & 782   & 11,403   & 7,069   & 2,183   & 151   & 73,420    \\ \cline{2-16} 
\multirow{-5}{*}{x64}                      & \multicolumn{14}{c?{1.1pt}}{Total \# Obfuscated Functions}  & \textbf{365,998}   \\ \midrule[1.1pt]
\multicolumn{15}{c?{1.1pt}}{Total \# Functions} & \textbf{1,472,066} \\ \bottomrule[1.1pt]

\end{tabular}
\end{table*}

\vspace{.1cm}\noindent\textbf{Micro-tracing.}
We implement micro-tracing by Unicorn~\cite{quynh2015unicorn}, a cross-platform CPU emulator based on QEMU~\cite{bellard2005qemu}. 
We micro-execute each function 3 times with different initialized registers and memory, generating 3 micro-traces (including both static code and dynamic values) for pretraining (masked LM).
We leverage multi-processing to parallelize micro-executing each function and set 30 seconds as the timeout for each run in case any instruction gets stuck (\ie infinite loops).
For each function (with 3 micro-traces), we append one additional dummy trace, which consists of only dummy values (\texttt{\#\#}). 
This setting encourages the model to leverage its learned execution semantics (from other traces with concrete dynamic values) to predict the masked code with \emph{only} code context, which helps the finetuning task as we only feed the functions' static code as input (discussed in Section~\ref{sec:intro}).

\vspace{.1cm}\noindent\textbf{Baselines.}
For comparing cross-architecture performance, we consider 2 state-of-the-art systems.
The first one is SAFE~\cite{massarelli2019safe} that achieves state-of-the-art function matching accuracy. 
As SAFE's trained model is publicly available, we compare \sys with both SAFE's reported results on their dataset, \ie OpenSSL-1.0.1f and OpenSSL-1.0.1u, and running their trained models on all of our collected binaries.
The second baseline is Gemini~\cite{xu2017neural}, which is shown outperformed by SAFE.
As Gemini does not release their trained models, we compare our results to their reported numbers directly on their evaluated dataset, \ie OpenSSL-1.0.1f and OpenSSL-1.0.1u.

For cross-optimization/obfuscation comparison, we consider Asm2Vec~\cite{ding2019asm2vec} and Blex~\cite{egele2014blanket} as the baselines.
Asm2Vec achieves the state-of-the-art cross-optimization/obfuscation results, based on learned embeddings from static assembly code.
Blex, on the other hand, leverages functions' dynamic behavior to match function binaries.
As we only find a third-party implementation of Asm2Vec that achieves extremely low Precision@1 (the metric used in Asm2Vec) from our testing (\eg 0.02 vs. their reported 0.814), and we have contacted the authors and do not get replies, we directly compare to their reported numbers. 
Blex is not publicly available either, so we also compare to their reported numbers directly.

\vspace{.1cm}\noindent\textbf{Metrics.}
As the cosine similarity between two function embeddings can be an arbitrary real value between -1 and 1, there has to be a mechanism to threshold the similarity score to determine whether the function pairs are similar or dissimilar.
The chosen thresholds largely determine the model's predictive performance~\cite{pagani2018beyond}.
To avoid the potential bias introduced by a specific threshold value, we consider the receiver operating characteristic (ROC) curve, which measures the model's false positives/true positives under different thresholds.
Notably, we use the area under curve (AUC) of the ROC curve to quantify the accuracy of \sys to facilitate benchmarking -- the higher the AUC score, the better the model's accuracy.

Certain baselines do not use AUC score to evaluate their system. For example, Asm2Vec uses Precision at Position 1 (Precision@1), and Blex uses the number of matched functions as the metric.
Therefore, we also include these metrics to evaluate \sys when needed.
Specifically, given a sequence of query functions and the target functions to be matched, Precision@1 measures the percentage of matched query functions in the target functions.
Here ``match'' means the query function should have the \emph{top-1 highest} similarity score with the ground truth function among the target functions.

To evaluate pretraining performance, we use the standard metric for evaluating the language model -- perplexity (PPL). The lower the PPL the better the pretrained model in predicting masked code.
The lowest PPL is 1.


\vspace{.1cm}\noindent\textbf{Pretraining setup.}
To strictly separate the functions in pretraining, finetuning, and testing, we pretrain \sys on all functions in the dataset, \emph{except the functions in the project going to be finetuned and evaluated for similarity.}
For example, the function matching results of Binutils (Table~\ref{tab:result}) are finetuned on the model pretrained on all other projects without Binutils.

Note that pretraining \emph{is agnostic to any labels} (\eg ground-truth indicating similar functions).
Therefore, we can always pretrain on large-scale codebases, which can potentially include the functions for finetuning (this is the common practice in transfer learning~\cite{devlin2018bert}).
It is thus worth noting that our setup of separating functions for pretraining and finetuning makes our testing significantly more challenging.

We keep the pretrained model weights that achieve the best validation PPL for finetuning.
The validation set for pretraining consists of 10,000 random functions selected from Table~\ref{tab:dataset}.

\vspace{.1cm}\noindent\textbf{Finetuning setup.}
We choose 50,000 random function pairs for each project and randomly select \emph{only 10\%} for training, and the remaining is used as the testing set.
We keep the training and testing functions \emph{strictly non-duplicated} by ensuring the functions that appear in training function pairs not appear in the testing. 
As opposed to the typical train-test split (\eg 80\% training and 20\% testing~\cite{massarelli2019safe}), our setting requires the model to generalize from few training samples to a large number of unseen testing samples, which alleviates the possibility of overfitting.
Moreover, we keep the ratio between similar and dissimilar function pairs in the finetuning set as roughly 1:5.
This setting follows the practice of contrastive learning~\cite{saunshi2019theoretical,chen2020simple}, respecting the actual distribution of similar/dissimilar functions as the number of dissimilar functions is often larger than that of similar functions in practice.

\vspace{.1cm}\noindent\textbf{Hyperparameters.}
We pretrain and finetune the models for 10 epochs and 30 epochs, respectively. We choose $\alpha=0.125$ in Equation~\ref{eq:pretrain} such that the cross-entropy loss of code prediction and value prediction have the same weight.
We pick $\xi=0.1$ in Equation~\ref{eq:cosembloss} to make the model slightly inclined to treat functions as dissimilar because functions in practice are mostly dissimilar.
We fix the largest input length to be 512 and split the functions longer than this length into subsequences for pretraining.
We average the subsequences' embeddings during finetuning if the function is split to more than one subsequences.
In this paper, we keep most of the hyperparameters fixed throughout the evaluation if not mentioned explicitly (complete hyperparameters are defined in Appendix~\ref{subsec:hyperparm}).
While we can always search for better hyperparameters, there is no principled method to date~\cite{bergstra2012random}.
We thus leave as future work a more thorough study of \sys's hyperparameters.

\section{Evaluation}
\label{sec:eval}

Our evaluation aims to answer the following questions.
\begin{itemize}
    \item RQ1: How accurate is \sys in matching semantically similar function binaries across different architectures, optimizations, and obfuscations?
    \item RQ2: How does \sys compare to the state-of-the-art?
    \item RQ3: How fast is \sys compared to other tools?
    \item RQ4: How much does pretraining on micro-traces help improve the accuracy of matching functions?
\end{itemize}





\begin{table}[!t]

\footnotesize
\setlength{\tabcolsep}{5.5pt}
\centering
\renewcommand{\arraystretch}{1}
\setlength\aboverulesep{0.4pt}
\setlength\belowrulesep{0.4pt}

\caption{\sys results (in AUC score) on function pairs across architectures, optimizations, and obfuscations. 
}
\label{tab:result}

\begin{tabular}{r?{1.1pt}ccccc}
\toprule[1.1pt]
\multirow{2}{*}{}    & \multicolumn{5}{c}{\textbf{Cross-}}                                      \\ \cline{2-6} 
 & \multicolumn{1}{c|}{\textbf{ARCH}} & \multicolumn{1}{c|}{\textbf{OPT}} & \multicolumn{1}{c|}{\textbf{OBF}} & \multicolumn{1}{c|}{\textbf{\begin{tabular}[c]{@{}l@{}}ARCH+\\ OPT\end{tabular}}} & \multicolumn{1}{c}{\textbf{\begin{tabular}[c]{@{}l@{}}ARCH+\\ OPT+\\ OBF\end{tabular}}} \\ \midrule[.9pt]
\rowcolor{lightblue} \textbf{Binutils}    & \multicolumn{1}{c|}{0.993} & \multicolumn{1}{c|}{0.993} & \multicolumn{1}{c|}{0.991} & \multicolumn{1}{c|}{0.959} & 0.947 \\ \hline
\textbf{Coreutils}   & \multicolumn{1}{c|}{0.991} & \multicolumn{1}{c|}{0.992} & \multicolumn{1}{c|}{0.991} & \multicolumn{1}{c|}{0.956} &  0.945 \\ \hline
\rowcolor{lightblue} \textbf{Curl}        & \multicolumn{1}{c|}{0.993} & \multicolumn{1}{c|}{0.993} & \multicolumn{1}{c|}{0.991} & \multicolumn{1}{c|}{0.958} & 0.956 \\ \hline
\textbf{Diffutils}   & \multicolumn{1}{c|}{0.992} & \multicolumn{1}{c|}{0.992} & \multicolumn{1}{c|}{0.990} & \multicolumn{1}{c|}{0.970} & 0.961 \\ \hline
\rowcolor{lightblue} \textbf{Findutils}   & \multicolumn{1}{c|}{0.990} & \multicolumn{1}{c|}{0.992} & \multicolumn{1}{c|}{0.990} & \multicolumn{1}{c|}{0.965} & 0.963 \\ \hline
\textbf{GMP}         & \multicolumn{1}{c|}{0.992} & \multicolumn{1}{c|}{0.993} & \multicolumn{1}{c|}{0.990} & \multicolumn{1}{c|}{0.968} & 0.966 \\ \hline
\rowcolor{lightblue} \textbf{ImageMagick} & \multicolumn{1}{c|}{0.993} & \multicolumn{1}{c|}{0.993} & \multicolumn{1}{c|}{0.989} & \multicolumn{1}{c|}{0.960} & 0.951 \\ \hline
\textbf{Libmicrohttpd}     & \multicolumn{1}{c|}{0.994} & \multicolumn{1}{c|}{0.994} & \multicolumn{1}{c|}{0.991} & \multicolumn{1}{c|}{0.972} & 0.969 \\ \hline
\rowcolor{lightblue} \textbf{LibTomCrypt} & \multicolumn{1}{c|}{0.992} & \multicolumn{1}{c|}{0.994} & \multicolumn{1}{c|}{0.991} & \multicolumn{1}{c|}{0.981} & 0.970 \\ \hline
\textbf{OpenSSL}     & \multicolumn{1}{c|}{0.992} & \multicolumn{1}{c|}{0.992} & \multicolumn{1}{c|}{0.989} & \multicolumn{1}{c|}{0.964} &  0.956 \\ \hline
\rowcolor{lightblue} \textbf{PuTTy}       & \multicolumn{1}{c|}{0.992} & \multicolumn{1}{c|}{0.995} & \multicolumn{1}{c|}{0.990} & \multicolumn{1}{c|}{0.961} & 0.952 \\ \hline
\textbf{SQLite}      & \multicolumn{1}{c|}{0.991} & \multicolumn{1}{c|}{0.994} & \multicolumn{1}{c|}{0.993} & \multicolumn{1}{c|}{0.980} & 0.959 \\ \hline
\rowcolor{lightblue} \textbf{Zlib}        & \multicolumn{1}{c|}{0.990} & \multicolumn{1}{c|}{0.991} & \multicolumn{1}{c|}{0.990} & \multicolumn{1}{c|}{0.979} & 0.965 \\ \midrule[.9pt]
\textbf{Average}     & \multicolumn{1}{c|}{0.992} & \multicolumn{1}{c|}{0.993} & \multicolumn{1}{c|}{0.990} & \multicolumn{1}{c|}{0.967} & 0.958 \\ \bottomrule[1.1pt]
\end{tabular}
\end{table}

\subsection{RQ1: Accuracy}
\label{subsec:rq1}

We evaluate how accurate \sys is in matching similar functions across different architectures, optimizations, and obfuscations. 
As shown in Table~\ref{tab:result}, we prepare function pairs for each project (first column) with 5 types of partitions.
(1) ARCH: the function pairs have \emph{different architectures} but same optimizations without obfuscations (2nd column).
(2) OPT: the function pairs have \emph{different optimizations} but same architectures without obfuscations (3rd column).
(3) OBF: the function pairs have \emph{different obfuscations} with same architectures (x64) and no optimization (4th column).
(4) ARCH+OPT: the function pairs have \emph{both different architectures and optimizations} without obfuscations (5th column).
(5) ARCH+OPT+OBF: the function pairs can come from arbitrary architectures, optimizations, and obfuscations (6th column).

Table~\ref{tab:result} reports the mean testing AUC scores of \sys on each project with 3 runs.
On average, \sys achieves $>0.958$ (and up to 0.995) AUC scores, even in the most challenging setting where the functions can come from different architectures, optimizations, and obfuscations at the same time. 
We note that \sys performs the best on cross-optimization matching. 
This is intuitive as the syntax of two functions from different optimizations are not changed significantly (\eg the name of opcode, operands remain the same).
Nevertheless, we find the AUC scores for matching functions from different architectures is only 0.001 lower, which indicates the model is robust to entirely different syntax between two architectures.
On matching functions with different obfuscations, \sys's results are 0.026, on average, lower than that of cross-optimizations, which indicates the obfuscation changes the code more drastically. 
Section~\ref{subsec:rq2} will show the specific results of \sys on each obfuscations.

\subsection{RQ2: Baseline Comparison}
\label{subsec:rq2}

\vspace{.1cm}\noindent\textbf{Cross-architecture search.} 
As described in Section~\ref{sec:impl}, we first compare \sys with SAFE and Gemini on OpenSSL-1.0.1f and OpenSSL-1.0.1u with their reported numbers (as they only evaluate on these two projects). 
We then run SAFE's released model on our dataset and compare to \sys.

We use our testing setup (see Section~\ref{sec:impl}) to evaluate SAFE's trained model, where 90\% of the total functions of those in Table~\ref{tab:dataset} are used to construct the testing set.
These testing sets are much larger than that in SAFE, where they only evaluate on 20\% of the OpenSSL functions.
Note that the dataset used in SAFE are all compiled by \texttt{GCC-5.4} at the time when it is publicized (November 2018), while ours are compiled by \texttt{GCC-7.5} (April 2020).
We find these factors (the more diverse dataset and different compilers) can all lead to the possible dataset distribution shift, which often results in the decaying performance of ML models when applied in the security applications~\cite{jordaney2017transcend}.

To study the distribution shift, we measure the KL-divergence~\cite{kullback1951information} between SAFE's dataset (OpenSSL compiled by \texttt{GCC-5.4}) and our dataset.
Specifically, we use the probability distribution of the raw bytes of the compiled projects' binaries, and compute their KL-divergence between SAFE and ours.
As OpenSSL is also a subset of our complete dataset, we compute the KL-divergence between our compiled OpenSSL and that of SAFE as the baseline.

We find the KL-divergence is 0.02 between SAFE's dataset and ours, while it decreases to 0.0066 between our compiled OpenSSL and that of SAFE.
This indicates that the underlying distribution of our test set shifts from that of SAFE's.
Moreover, the KL-divergence of 0.0066 between the same dataset (OpenSSL) but only compiled by different GCC versions implies that the compiler version has much smaller effect to the distribution shift than the different software projects.

\begin{figure}[!t]
\centering

\includegraphics[width=0.8\linewidth]{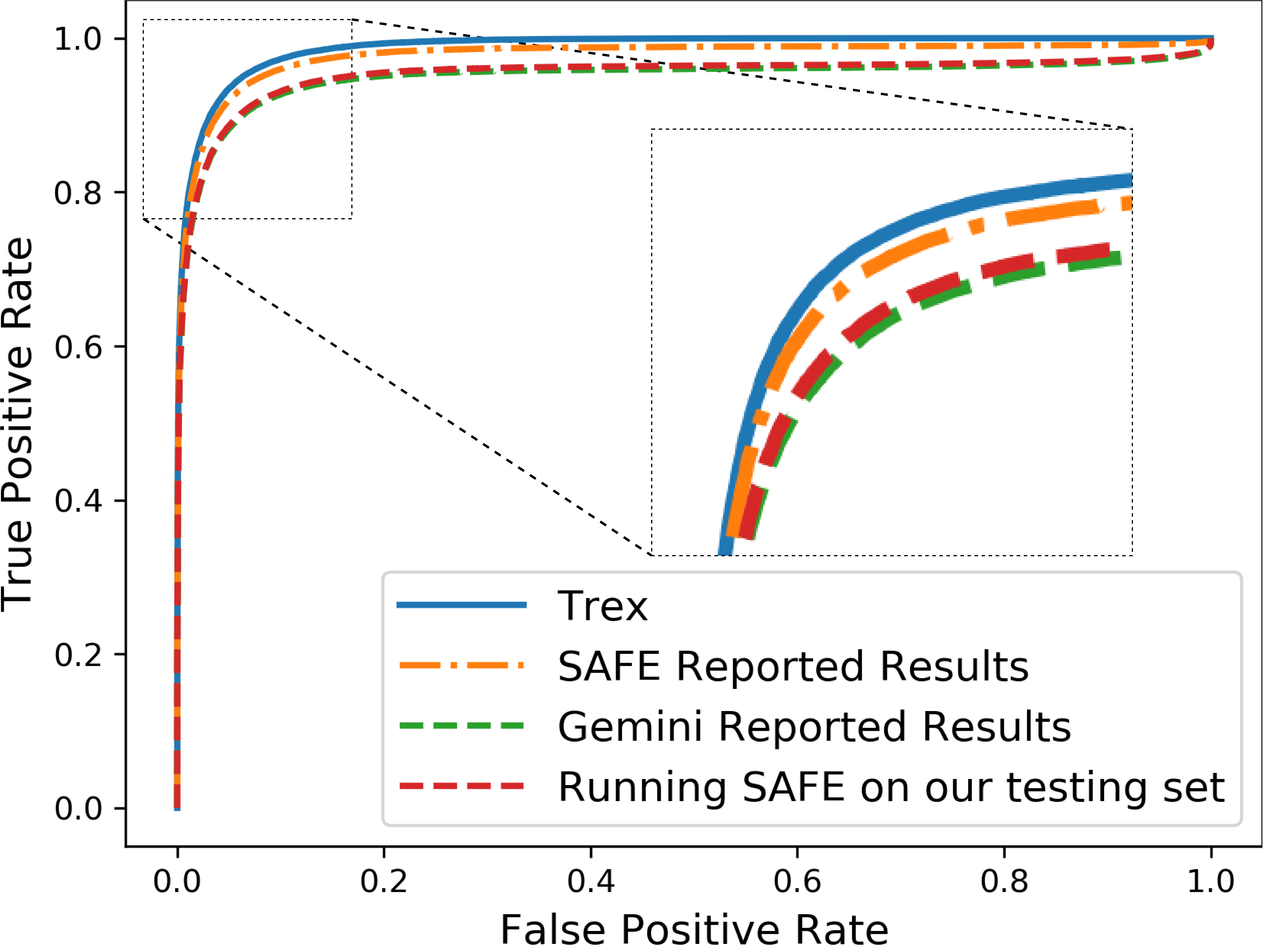}

\caption{ROC curves of matching functions in OpenSSL across different architectures. \sys outperforms the reported results of SAFE and Gemini and the results of running SAFE's trained model on our testing set.}
\label{fig:cross-arch-roc}
\end{figure}

As shown in Figure~\ref{fig:cross-arch-roc}, our ROC curve is higher than those reported in SAFE and Gemini. 
While SAFE's reported AUC score, 0.992, is close to ours, when we run their trained model on our testing set, its AUC score drops to 0.976 -- possibly because our testing set is much larger than theirs.
This observation demonstrates the generalizability of \sys -- when pretrained to approximately learn execution semantics explicitly, it can quickly generalize to match unseen (semantically similar) functions with only a minimal training set.

\begin{figure}[!t]
\centering

\includegraphics[scale=.4]{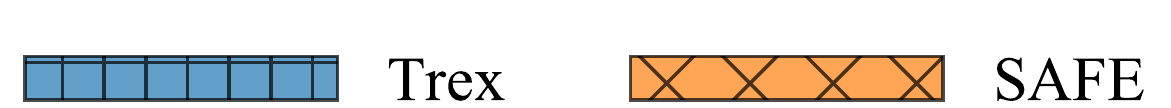}
\vspace{.1cm}
\includegraphics[width=\linewidth]{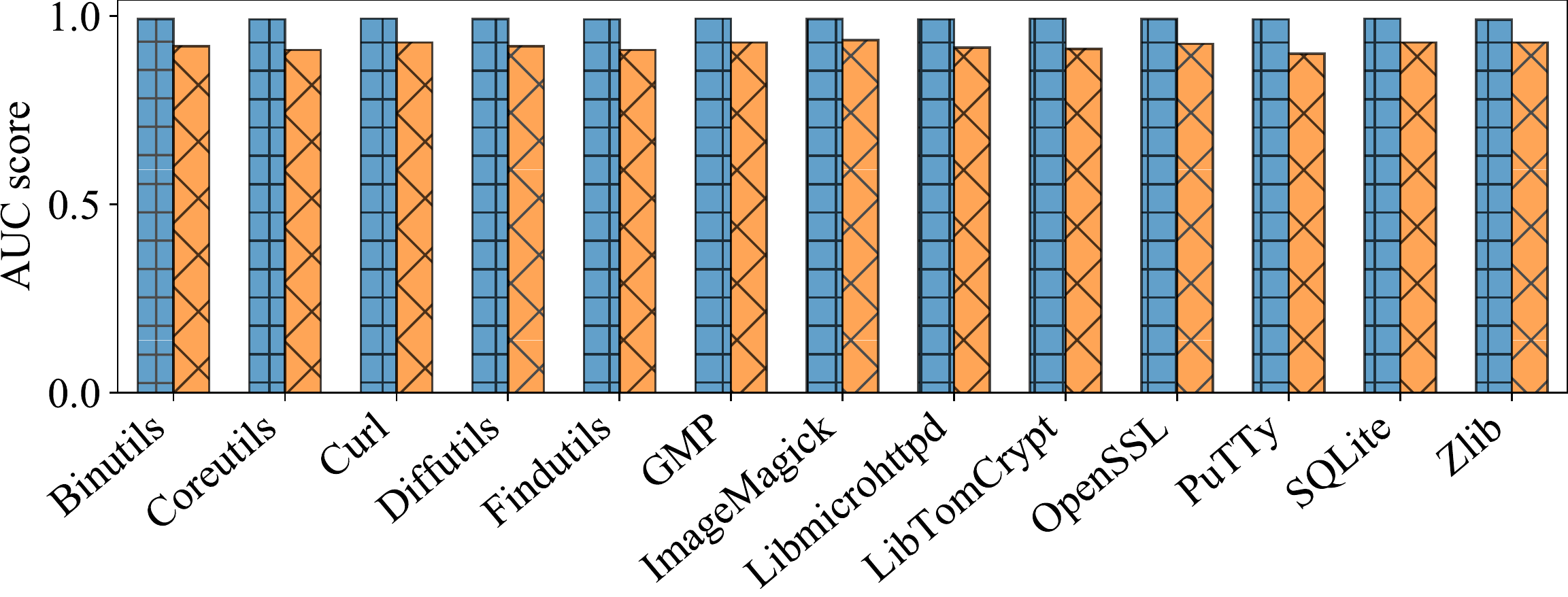}

\caption{Comparison between \sys and SAFE on matching functions in each project compiled to different architectures (see Table~\ref{tab:dataset}).}
\label{fig:cross-arch-auc-safe}
\end{figure}

Figure~\ref{fig:cross-arch-auc-safe} shows that \sys consistently outperforms SAFE on all projects, \ie by 7.3\% on average.
As the SAFE's model is only trained on OpenSSL, we also follow the same setting by training \sys on only OpenSSL, similar to the cross-project setting described in Section~\ref{subsec:rq1}.

\vspace{.1cm}\noindent\textbf{Cross-optimization search.}
We compare \sys with Asm2Vec and BLEX on matching functions compiled by different optimizations.
As both Asm2vec and Blex run on single architecture, we restrict the comparison on x64.
Besides, since Asm2Vec uses Precision@1 and Blex uses accuracy as the metric (discussed in Section~\ref{sec:impl}), we compare with each tool separately using their metrics and on their evaluated dataset.




\begin{table}[!t]

\footnotesize
\setlength{\tabcolsep}{8pt}
\centering
\renewcommand{\arraystretch}{1}
\setlength\aboverulesep{0.4pt}
\setlength\belowrulesep{0.4pt}

\caption{Comparison between \sys and Asm2Vec (in Precision@1) on function pairs across optimizations.
}
\label{tab:result-asm2vec}

\begin{tabular}{r?{1.1pt}cccc}
\toprule[1.1pt]
\multicolumn{1}{c?{1.1pt}}{} & \multicolumn{4}{c}{Cross Compiler Optimization}                                                                \\ \cline{2-5} 
\multicolumn{1}{l?{1.1pt}}{} & \multicolumn{2}{c?{1.1pt}}{\texttt{O2} and \texttt{O3}}                                   & \multicolumn{2}{c}{\texttt{O0} and \texttt{O3}}               \\ \cline{2-5} 
\multicolumn{1}{c?{1.1pt}}{\textbf{}} & \multicolumn{1}{c|}{\sys} & \multicolumn{1}{c?{1.1pt}}{Asm2Vec} & \multicolumn{1}{c|}{\sys} & Asm2Vec \\ \midrule[.9pt]
\rowcolor{lightblue} Coreutils             & \multicolumn{1}{c|}{\textbf{0.955}} & \multicolumn{1}{c?{1.1pt}}{0.929} & \multicolumn{1}{c|}{\textbf{0.913}} & 0.781 \\ \hline
Curl                  & \multicolumn{1}{c|}{\textbf{0.961}} & \multicolumn{1}{c?{1.1pt}}{0.951} & \multicolumn{1}{c|}{\textbf{0.894}} & 0.850 \\ \hline
\rowcolor{lightblue} GMP                   & \multicolumn{1}{c|}{\textbf{0.974}} & \multicolumn{1}{c?{1.1pt}}{0.973} & \multicolumn{1}{c|}{\textbf{0.886}} & 0.763 \\ \hline
ImageMagick           & \multicolumn{1}{c|}{\textbf{0.971}} & \multicolumn{1}{c?{1.1pt}}{0.971} & \multicolumn{1}{c|}{\textbf{0.891}} & 0.837 \\ \hline
\rowcolor{lightblue} LibTomCrypt           & \multicolumn{1}{c|}{\textbf{0.991}} & \multicolumn{1}{c?{1.1pt}}{0.991} & \multicolumn{1}{c|}{\textbf{0.923}} & 0.921 \\ \hline
OpenSSL               & \multicolumn{1}{c|}{\textbf{0.982}} & \multicolumn{1}{c?{1.1pt}}{0.931} & \multicolumn{1}{c|}{\textbf{0.914}} & 0.792 \\ \hline
\rowcolor{lightblue} PuTTy                 & \multicolumn{1}{c|}{\textbf{0.956}} & \multicolumn{1}{c?{1.1pt}}{0.891} & \multicolumn{1}{c|}{\textbf{0.926}} & 0.788 \\ \hline
SQLite                & \multicolumn{1}{c|}{\textbf{0.931}} & \multicolumn{1}{c?{1.1pt}}{0.926} & \multicolumn{1}{c|}{\textbf{0.911}} & 0.776 \\ \hline
\rowcolor{lightblue} Zlib                  & \multicolumn{1}{c|}{\textbf{0.890}} & \multicolumn{1}{c?{1.1pt}}{0.885} & \multicolumn{1}{c|}{\textbf{0.902}} & 0.722 \\ \midrule[.9pt]
Average               & \multicolumn{1}{c|}{\textbf{0.957}} & \multicolumn{1}{c?{1.1pt}}{0.939} & \multicolumn{1}{c|}{\textbf{0.907}} & 0.803 \\ \bottomrule[1.1pt]
\end{tabular}

\end{table}

Table~\ref{tab:result-asm2vec} shows \sys outperforms Asm2Vec in Precision@1 (by 7.2\% on average) on functions compiled by different optimizations (\ie between \texttt{O2} and \texttt{O3} and between \texttt{O0} and \texttt{O3}).
As the syntactic difference introduced by optimizations between \texttt{O0} and \texttt{O3} is more significant than that between \texttt{O2} and \texttt{O3}, both tools have certain level of decrease in AUC scores (5\% drop for \sys and 14\% for Asm2Vec), but \sys's AUC score drops much less than that of Asm2Vec.

To compare to Blex, we evaluate \sys on Coreutils between optimizations \texttt{O0} and \texttt{O3}, where they report to achieve better performance than BinDiff~\cite{bindiff}.
As Blex show the matched functions of each individual utility in Coreutils in a barchart without including the concrete numbers of matched functions, we estimate their matched functions using their reported average percentage (75\%) on all utilities.

\begin{figure}[!t]
\centering

\includegraphics[width=.95\linewidth]{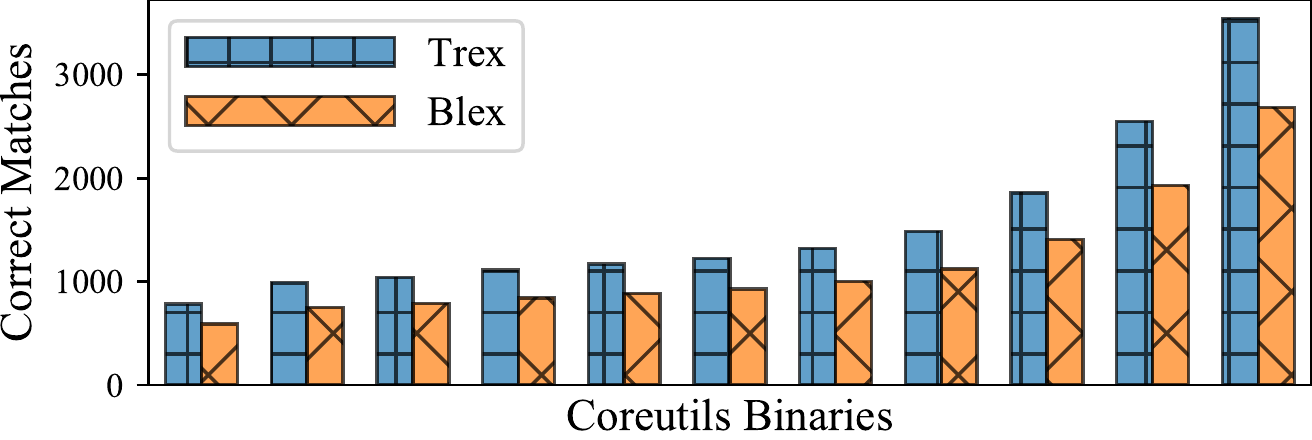}

\caption{Cross-optimization function matching between \texttt{O0} and \texttt{O3} on Coreutils by \sys and Blex. We sort the 109 utility binaries in Coreutils by their number of functions, and aggregate the matched functions every 10 utilities.}
\label{fig:cross-opt-blex}
\end{figure}

Figure~\ref{fig:cross-opt-blex} shows that \sys consistently outperforms Blex in number of matched functions in all utility programs of Coreutils.
Note that Blex also executes the function and uses the dynamic features to match binaries.
The observation here thus implies that the learned execution semantics from \sys is more effective than the hand-coded features in Blex for matching similar binaries.

\vspace{.1cm}\noindent\textbf{Cross-obfuscation search.}
We compare \sys to Asm2Vec on matching obfuscated function binaries with different obfuscation methods.
Notably, Asm2Vec is evaluated on obfuscations including bogus control flow (\texttt{bcf}), control flow flattening (\texttt{cff}), and instruction substitution (\texttt{sub}), which are subset of our evaluated obfuscations (Table~\ref{tab:dataset}).
As Asm2Vec only evaluates on 4 projects, \ie GMP, ImageMagic, LibTomCrypt, and OpenSSL, we focus on these 4 projects and \sys's results for other projects are included in Table~\ref{tab:result}.

\begin{table}[!t]

\footnotesize
\setlength{\tabcolsep}{2.5pt}
\centering
\renewcommand{\arraystretch}{1}
\setlength\aboverulesep{0.4pt}
\setlength\belowrulesep{0.4pt}

\caption{Comparison between \sys and Asm2Vec (in Precision@1) on function pairs across differet obfuscations.
}
\label{tab:cross-obf-asm2vec}

\begin{tabular}{r|r?{1.1pt}l|l|l|l?{1.1pt}l}
\toprule[1.1pt]
                              &         & GMP   & LibTomCrypt & ImageMagic & OpenSSL & Average \\ \midrule[.9pt]
\rowcolor{lightblue}  & \sys    & \textbf{0.926} & \textbf{0.938}       & \textbf{0.934}      & \textbf{0.898}   & \textbf{0.924}   \\ \cline{2-7} 
\rowcolor{lightblue} \multirow{-2}{*}{\texttt{bcf}}                              & Asm2Vec & 0.802 & 0.920       & 0.933      & 0.883   & 0.885   \\ \midrule[.9pt]
\multirow{2}{*}{\texttt{ccf}} & \sys    & \textbf{0.943} & \textbf{0.931}       & \textbf{0.936}      & \textbf{0.940}   & \textbf{0.930}   \\ \cline{2-7} 
                              & Asm2Vec & 0.772 & 0.920       & 0.890      & 0.795   & 0.844   \\ \midrule[.9pt]
\rowcolor{lightblue} & \sys    & \textbf{0.949} & \textbf{0.962}       & \textbf{0.981}       & \textbf{0.980}   & \textbf{0.968}   \\ \cline{2-7} 
\rowcolor{lightblue} \multirow{-2}{*}{\texttt{sub}}                              & Asm2Vec & 0.940 & 0.960       & 0.981      & 0.961   & 0.961   \\ \midrule[.9pt]
\multirow{2}{*}{All}          & \sys    & \textbf{0.911} & \textbf{0.938}       & \textbf{0.960}      & \textbf{0.912}   & \textbf{0.930}   \\ \cline{2-7} 
                              & Asm2Vec & 0.854 & 0.880       & 0.830      & 0.690   & 0.814   \\ \bottomrule[1.1pt]
\end{tabular}
\end{table}

Table~\ref{tab:cross-obf-asm2vec} shows \sys achieves better Precision@1 score (by 14.3\% on average) throughout all different obfuscations. 
Importantly, the last two rows show when multiple obfuscations are combined, \sys performance is not dropping as significant as Asm2Vec.
It also shows \sys remains robust under varying obfuscations with different difficulties. 
For example, instruction substitution simply replaces very limited instructions (\ie arithmetic operations as shown in Section~\ref{sec:overview}) while control flow flattening dramatically changes the function code.
Asm2Vec has 12.2\% decreased score when the obfuscation is changed from \texttt{sub} to \texttt{ccf}, while \sys only decreases by 4\%.

\subsection{RQ3: Execution Time}
\label{subsec:rq3}

We evaluate the runtime performance of generating function embeddings for computing similarity.
We compare \sys with SAFE and Gemini on generating functions in 4 projects on x64 compiled by \texttt{O3}, \ie Binutils, Putty, Findutils, and Diffutils, which have disparate total number of functions (see Table~\ref{tab:dataset}.
This tests how \sys scales to different number of functions compared to other baselines.
Since the offline training (\ie pretraining \sys) of all the learning-based tools is a one-time cost, it can be amortized in the function matching process so we do not explicitly measure the training time.
Moreover, the output of all tools are function embeddings, which can be indexed and efficiently searched using locality sensitive hashing (LSH)~\cite{gionis1999similarity, rajaraman2011mining}.
Therefore, we do not compare the matching time of function embeddings as it simply depends on the performance of underlying LSH implementation.

Particularly, we compare the runtime of two procedures in matching functions.
(1) Function parsing, which transforms the function binaries into the format that the model needs. 
(2) Embedding generation, which takes the parsed function binary as input and computes function embedding. We test the embedding generation using our GPU (see Section~\ref{sec:impl}).

\begin{figure}[!t]
\centering

\includegraphics[scale=.4]{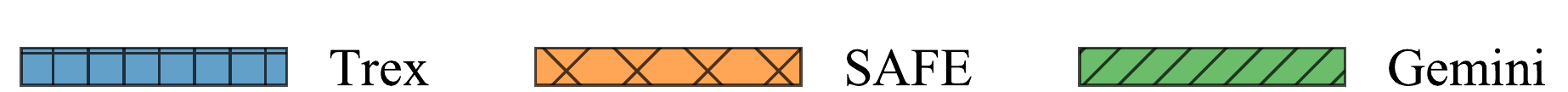}
\subfloat[Function parsing]{
\includegraphics[width=0.47\linewidth]{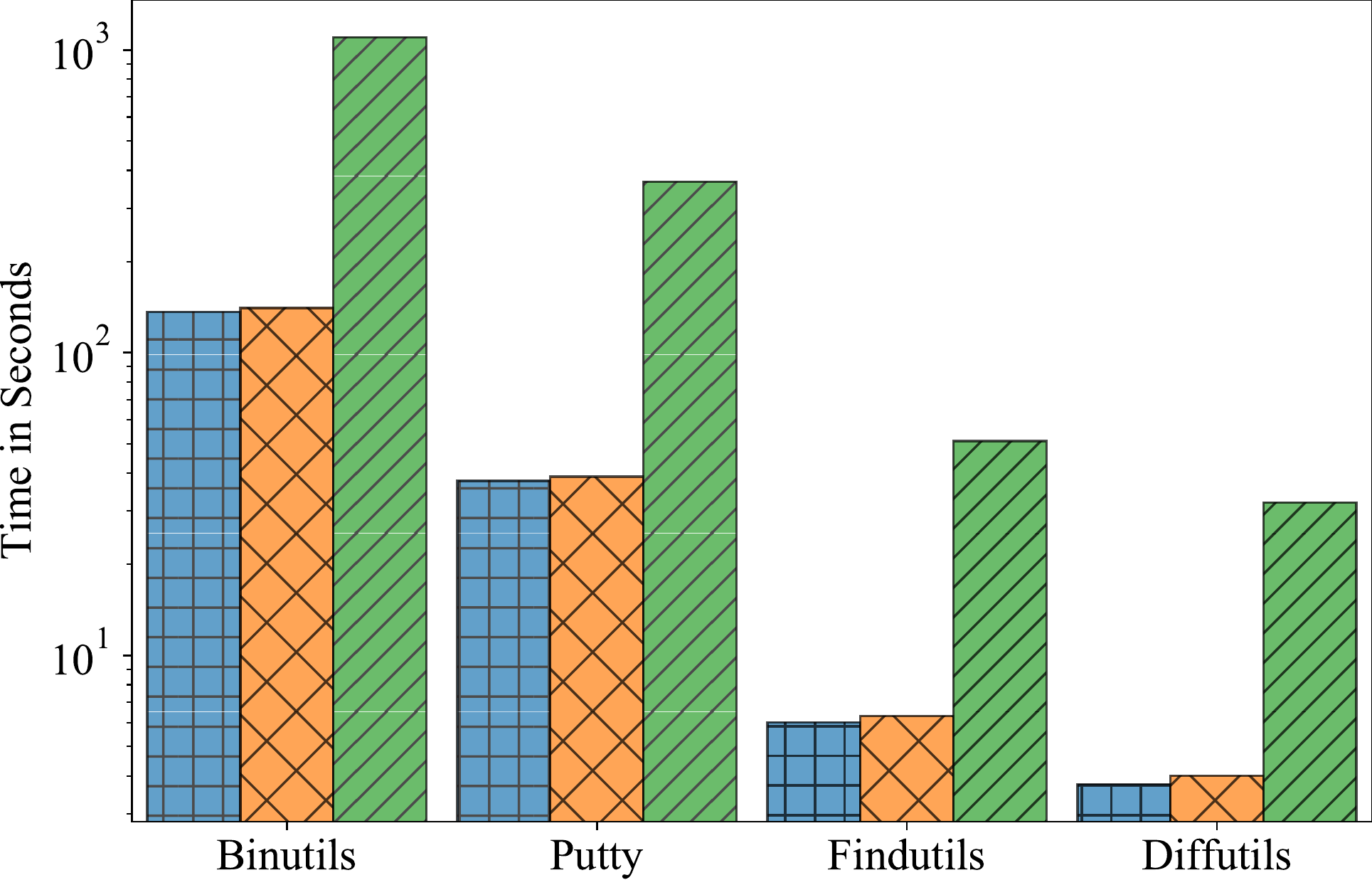}
\label{subfig:runtime_bar_parse}}
\subfloat[Embedding generation]{
\includegraphics[width=0.47\linewidth]{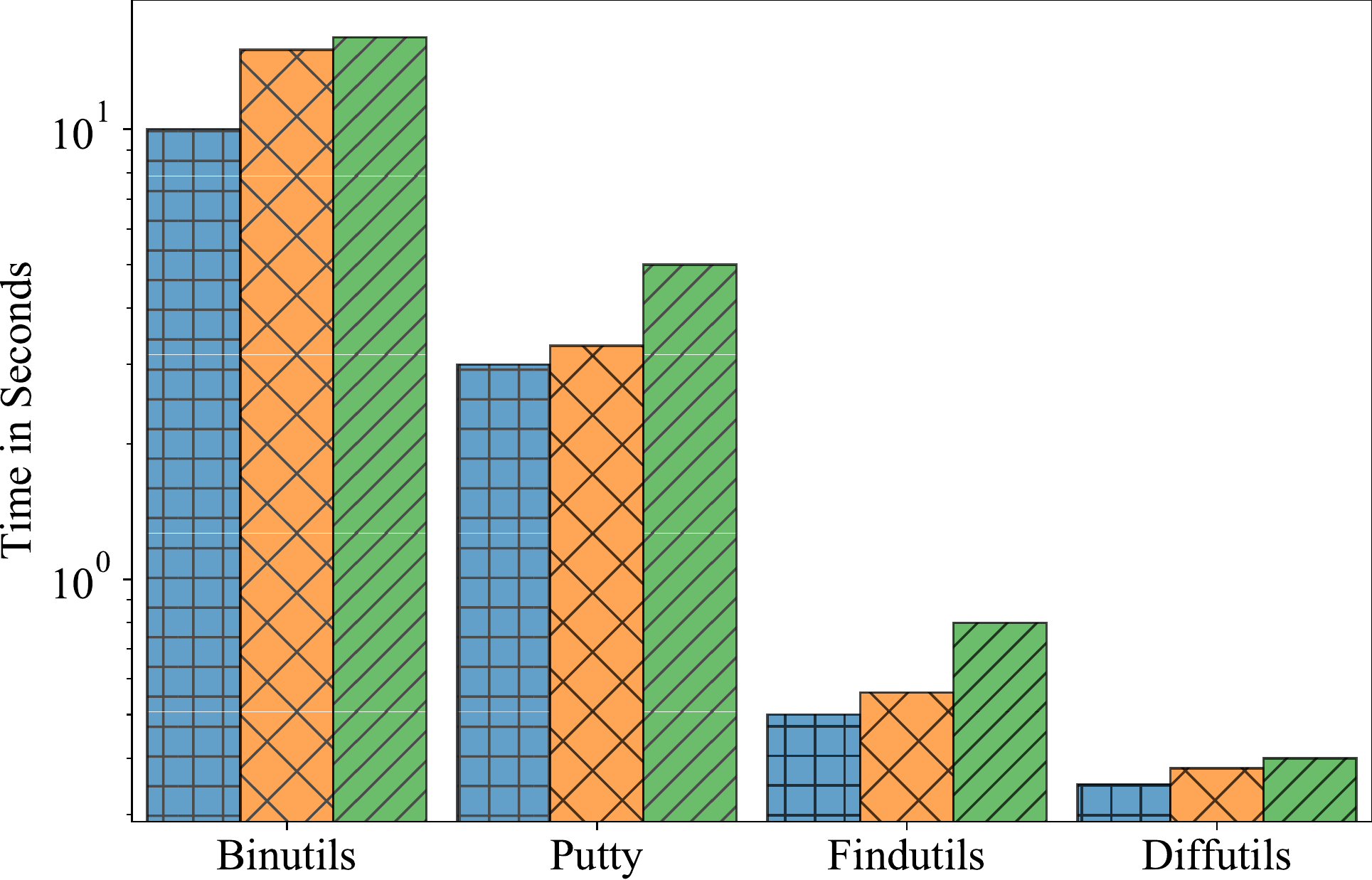}
\label{subfig:runtime_bar_emb}}

\caption{Runtime performance (lower is better) of \sys, SAFE, and Gemini on (a) function parsing and (b) embedding generation. The time is log-scaled.}
\label{fig:runtime}
\end{figure}

Figure~\ref{fig:runtime} shows that \sys is more efficient than the other tools in both function parsing and embedding generation for functions from 4 different projects with different number of functions (Table~\ref{tab:dataset}).
Gemini requires manually constructing control flow graph and extracting inter-/intra-basic-block feature engineering. It thus incurs the largest overhead.
For generating function embeddings, our underlying network architectures leverage Transformer self-attention layers, which is more amenable to parallezation with GPU than the recurrent (used by SAFE) and graph neural network (used by Gemini)~\cite{vaswani2017attention}.
As a result, \sys runs up to 8$\times$ faster than SAFE and Gemini.


\subsection{RQ4: Ablation Study}
\label{subsec:rq4}

In this section, we aim to quantify how much each key component in \sys's design helps the end results.
We first study how much does pretraining, argued to assist learning approximate execution semantics, help match function binaries.
We then study how does pretraining \emph{without micro-traces} affect the end results.
We also test how much does incorporating the micro-traces in the pretraining tasks improve the accuracy.

\vspace{.1cm}\noindent\textbf{Pretraining effectiveness.}
We compare the testing between AUC scores achieved by \sys (1) with pretraining (except the target project that will be finetuned), (2) with 66\% of pretraining functions in (1), (3) with 33\% of pretraining functions in (1), and (4) without pretraining (the function embedding is computed by randomly-initialized model weights that are not pretrained). 
The function pairs can come from arbitrary architectures, optimizations, and obfuscations.

\begin{figure}[!t]
\centering

\includegraphics[scale=.4]{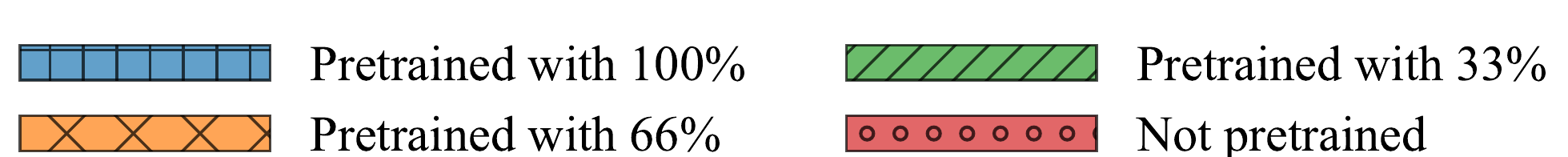}

\includegraphics[width=.95\linewidth]{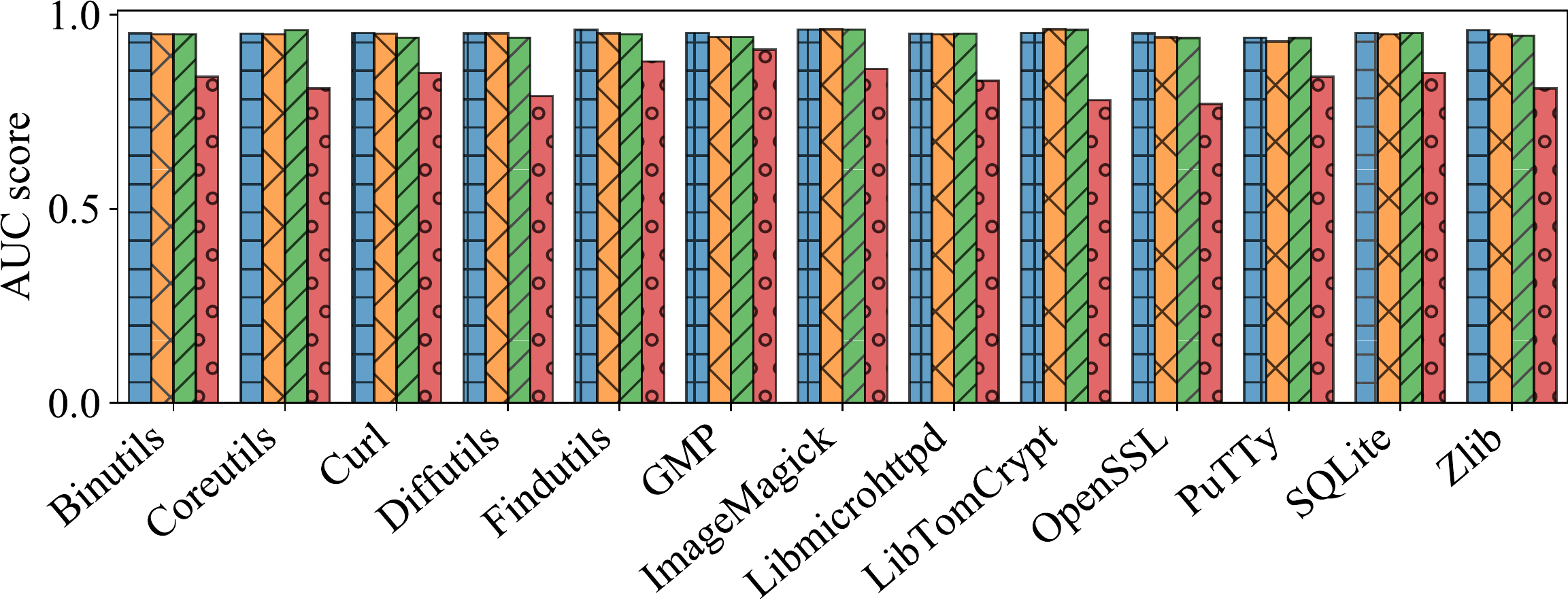}

\caption{Comparison of testing AUC scores between models pretrained with different fraction of the pretraining set.}
\label{fig:ablation-pretrain}
\end{figure}

Figure~\ref{fig:ablation-pretrain} shows that the model's AUC score drops significantly (on average 15.7\%) when the model is not pretrained.
Interestingly, we observe that the finetuned models achieve similar AUC scores, \ie with only 1\% decrease when pretrained with 33\% of the functions compared to pretraining with all functions.
This indicates that we can potentially decrease our pretraining data size to achieve roughly the same performance.
However, since our pretraining task does not require collect any label, we can still collect unlimited binary code found in the wild to enrich the semantics that can be observed.

\vspace{.1cm}\noindent\textbf{Pretraining w/o micro-traces.} 
We try to understand whether including the micro-traces in pretraining can really help the model to learn better execution semantics than learning from only static assembly code, which in turn results in better function matching accuracy.
Specifically, we pretrain the model on the data that contains only dummy value sequence (see Section~\ref{sec:method}), and follow the same experiment setting as described above.
Besides replacing the input value sequence as dummy value, we accordingly remove the prediction of dynamic values in the pretraining objective (Equation~\ref{eq:pretrain}).
We also compare to SAFE in this case.

Figure~\ref{fig:ablation-pretrain-finetune-trace} shows that the AUC scores decrease by 7.2\% when the model is pretrained without micro-trace.
The AUC score of pretrained \sys without micro-traces is even 0.035 lower than that of SAFE.
However, the model still performs reasonably well, achieving 0.88 AUC scores even when the functions can come from arbitrary architectures, optimizations, and obfuscations.
Moreover, we observe that pretraining without micro-traces has less performance drop than the model simply not pretrained (7.2\% vs. 15.7\%).
This demonstrates that even pretraining with only static assembly code is indeed helpful to improve matching functions.
One possible interpretation is that similar functions are statically similar in syntax, while understanding their inherently similar execution semantics just further increases the similarity score.

\begin{figure}[!t]
\centering

\includegraphics[scale=.3]{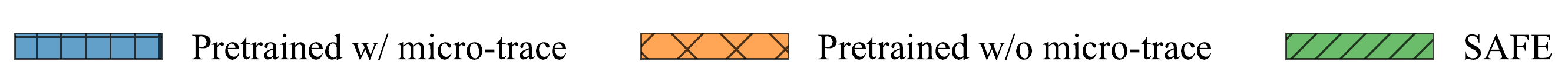}

\includegraphics[width=\linewidth]{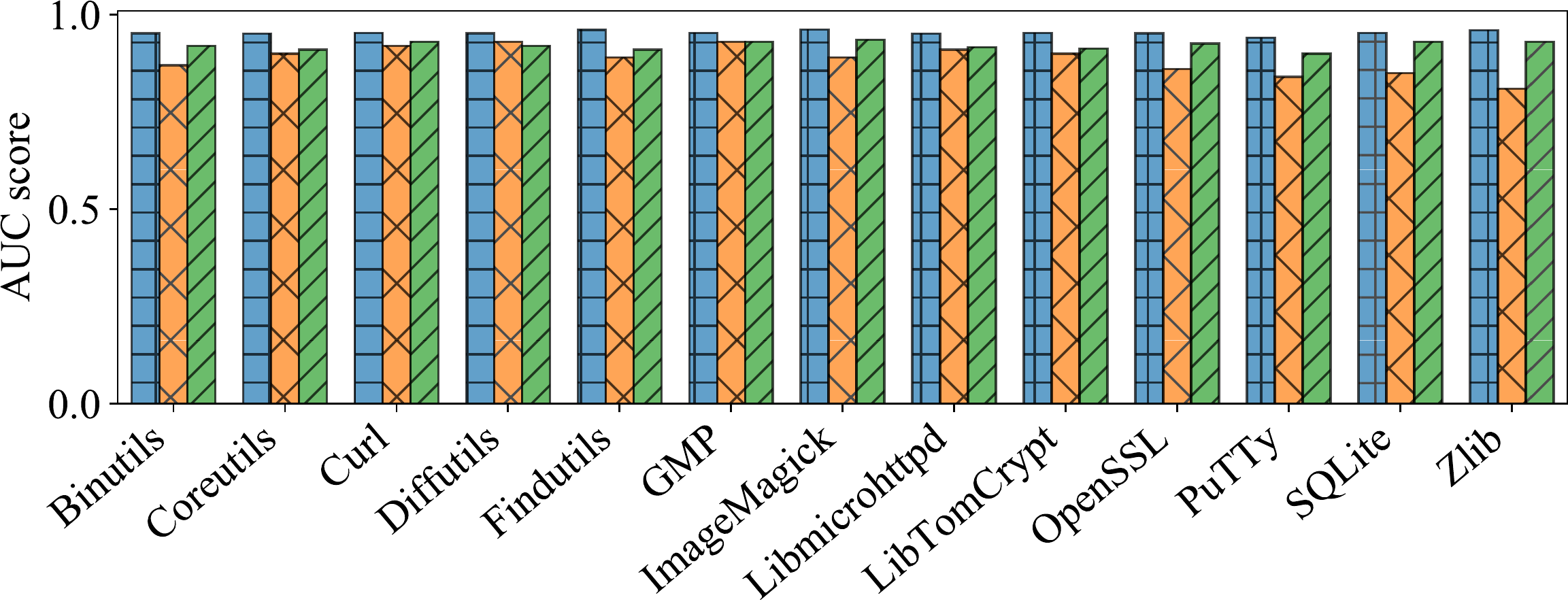}

\caption{Comparison of testing AUC scores between models pretrained \emph{with} micro-trace and pretrained \emph{without} micro-trace.}
\label{fig:ablation-pretrain-finetune-trace}
\end{figure}




\vspace{.1cm}\noindent\textbf{Pretraining accuracy.}
We study the testing perplexity (PPL defined in Section~\ref{sec:impl}) of pretraining \sys to directly validate whether it indeed helps \sys to learn the approximate execution semantics.
The rationale is the good performance, \ie the low PPL, on unseen testing function binaries indicates that \sys highly likely learns to generalize based on its learned approximate execution semantics.

\begin{figure}[!t]
\centering

\includegraphics[width=0.85\linewidth]{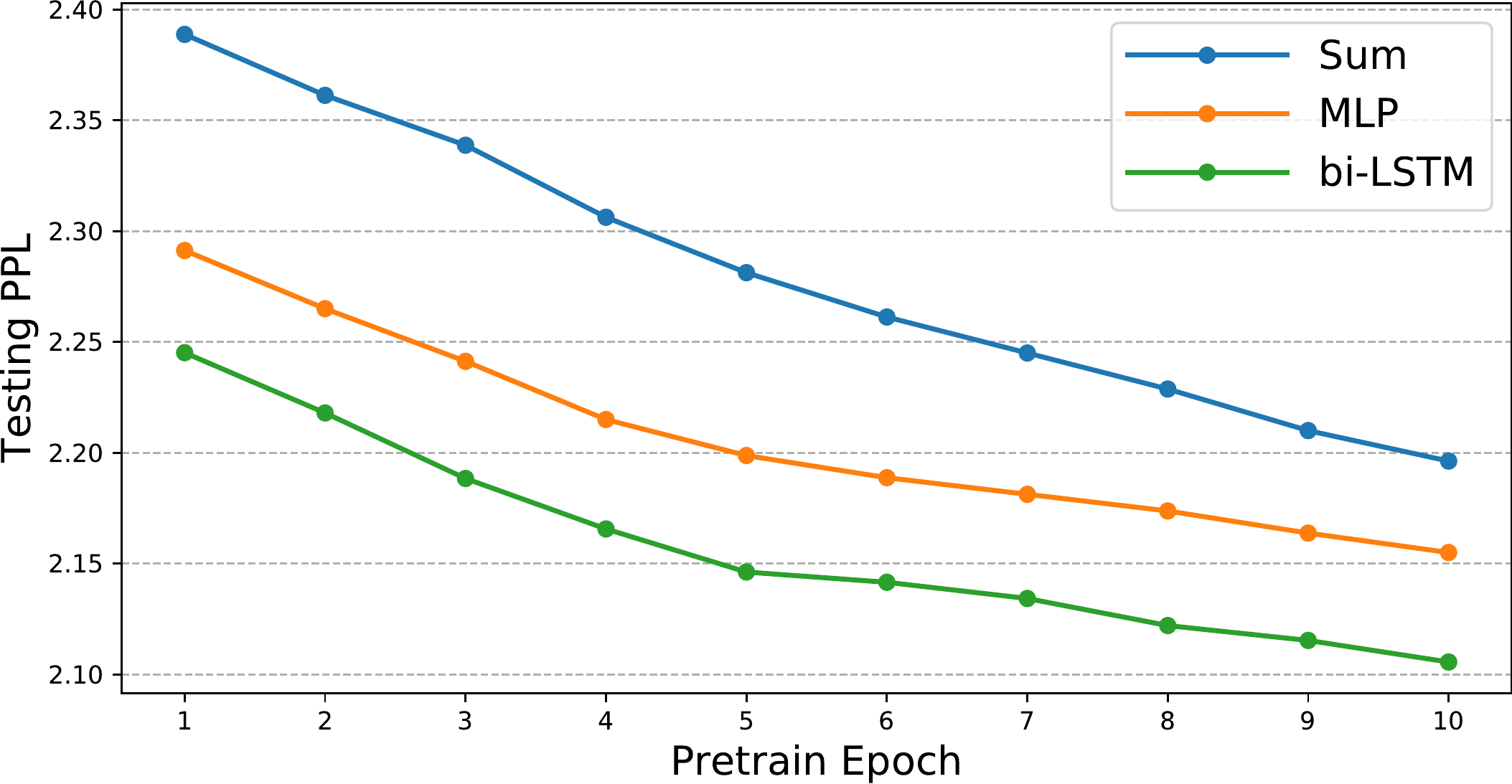}

\caption{Testing PPL of pretraining \sys in 10 epochs. We compare different designs of combining byte-sequence in the micro-trace (see Section~\ref{subsec:input_repr}).}
\label{fig:pretrain-loss}
\end{figure}

The green line in Figure~\ref{fig:pretrain-loss} shows the validation PPL in 10 epochs of pretraining \sys. 
The testing set is constructed by sampling 10,000 random functions from the projects used in pretraining (as described in Section~\ref{sec:impl}).
We observe that the PPL of \sys drops to close to 2.1, which is far below that of random guessing (\eg random guessing PPL is $2^{-\log(1/256)}=256$), indicating that \sys has around 0.48 confidence on average on the masked tokens being the correct value.
Note that a random guessing has only 1/256=0.004 confidence on the masked tokens being the correct value.



\section{Case Studies}
\label{sec:case}

In this section, we study how \sys can help discover new vulnerabilities in a large volume of latest firmware images.


\begin{table}[!t]

\footnotesize
\setlength{\tabcolsep}{4pt}
\centering
\renewcommand{\arraystretch}{1}
\setlength\aboverulesep{0.4pt}
\setlength\belowrulesep{0.4pt}

\caption{Vulnerabilities we have confirmed (\cmark) in firmware images (latest version) from 4 well-known vendors and products.}
\label{tab:cve_case}

\begin{tabular}{r?{1.1pt}c|c|c|c}
\toprule[1.1pt]
\multicolumn{1}{c|}{\multirow{2}{*}{\textbf{CVE}}} &
  \multicolumn{1}{c|}{\multirow{2}{*}{\textbf{\begin{tabular}[c]{@{}c@{}}Ubiquiti\\ sunMax\end{tabular}}}} &
  \multicolumn{1}{c|}{\multirow{2}{*}{\textbf{\begin{tabular}[c]{@{}c@{}}TP-Link\\ Deco-M4\end{tabular}}}} &
  \multicolumn{1}{c|}{\multirow{2}{*}{\textbf{\begin{tabular}[c]{@{}c@{}}NETGEAR\\ R7000\end{tabular}}}} &
  \multicolumn{1}{c}{\multirow{2}{*}{\textbf{\begin{tabular}[c]{@{}c@{}}Linksys\\ RE7000\end{tabular}}}} \\
\multicolumn{1}{c|}{} & \multicolumn{1}{c|}{} & \multicolumn{1}{c|}{} & \multicolumn{1}{c|}{} & \multicolumn{1}{c}{} \\ \midrule[.9pt]
CVE-2019-1563  & \cmark   & \cmark & \cmark &    \xmark                    \\ \hline
\rowcolor{lightblue} CVE-2017-16544 & \xmark   & \cmark & \xmark &  \xmark               \\ \hline
CVE-2016-6303  & \cmark   & \cmark & \cmark &   \cmark                                     \\ \hline
\rowcolor{lightblue} CVE-2016-6302  & \cmark   & \cmark & \cmark &  \cmark                     \\ \hline
CVE-2016-2842  & \cmark   & \cmark & \cmark &    \cmark                    \\ \hline
\rowcolor{lightblue} CVE-2016-2182  & \cmark   & \cmark & \cmark &  \cmark                     \\ \hline
CVE-2016-2180  & \cmark   & \cmark & \cmark &   \cmark                          \\ \hline
\rowcolor{lightblue} CVE-2016-2178  & \cmark   & \cmark & \cmark &  \cmark                         \\ \hline
CVE-2016-2176  & \cmark   & \cmark & \xmark &   \cmark                            \\ \hline
\rowcolor{lightblue} CVE-2016-2109  & \cmark   & \cmark & \xmark &  \cmark                     \\ \hline
CVE-2016-2106  & \cmark   & \cmark & \xmark &   \cmark                            \\ \hline
\rowcolor{lightblue} CVE-2016-2105  & \cmark   & \cmark & \xmark &   \cmark                            \\ \hline
CVE-2016-0799  & \cmark   & \cmark & \xmark &   \cmark                          \\ \hline
\rowcolor{lightblue} CVE-2016-0798  & \cmark   & \cmark & \xmark &  \cmark                     \\ \hline
CVE-2016-0797  & \cmark   & \cmark & \xmark &  \cmark                     \\ \hline
\rowcolor{lightblue} CVE-2016-0705  & \cmark   & \cmark & \xmark &  \cmark                            \\ \bottomrule[1.1pt]
\end{tabular}

\end{table}

Firmware images often include third-party libraries (\ie BusyBox, OpenSSL).
However, these libraries are frequently patched, but the manufacturers often fall behind to update them accordingly~\cite{OWASP}. 
Therefore, we study whether our tool can uncover function binaries in firmware images similar to known vulnerable functions.
We find existing state-of-the-art binary similarity tools (\ie SAFE~\cite{massarelli2019safe}, Asm2Vec~\cite{ding2019asm2vec}, Gemini~\cite{xu2017neural}) all perform their case studies on the firmware images and vulnerabilities that have already been studied before them~\cite{costin2014large, feng2016scalable, david2016statistical, chen2016towards}.
Therefore, we decide to collect our own dataset with more updated firmware images and the latest vulnerabilities, instead of reusing the existing benchmarks.
This facilitates finding new (1-day) vulnerabilities in most recent firmware images that are not disclosed before.

We crawl firmware images (in their latest version) in 180 products including WLAN routers, smart cameras, and solar panels, from well-known manufacturers' official releases and third-party providers such as DD-WRT~\cite{ddwrt}, as shown in Table~\ref{tab:firmware}.
We collect firmware images with only the latest version, which are much more updated than those studied in the state-of-the-art (dated before 2015, which are all patched before their studies).
For each function in the firmware images, we construct function embedding and build a firmware image database using Open Distro for Elasticsearch~\cite{elastic}, which supports vector-based indexing with efficient search support based on NMSLIB~\cite{boytsov2013engineering}.

We extract the firmware images using binwalk~\cite{binwalk}. 
In total, we collect 180 number of firmware images from 22 vendors.
These firmware images are developed by the original vendors, or from third-party providers such as DD-WRT~\cite{ddwrt}.
Most of them are for WLAN routers, while some are deployed in other embedded systems, such as solar panels and smart cameras.
Among the firmware images, 127 of them are compiled for MIPS 32-bit and 53 are compiled for ARM 32-bit.
169 of them are 2020 models, and the rest are 2019 and 2018 models.

\begin{table}[!t]

\footnotesize
\setlength{\tabcolsep}{7pt}
\centering
\renewcommand{\arraystretch}{1}
\setlength\aboverulesep{0.4pt}
\setlength\belowrulesep{0.4pt}

\caption{We study 16 vulnerabilities from OpenSSL and BusyBox, two widely-used libraries in firmware images.}
\label{tab:cve}

\begin{tabular}{r|l|l}
\toprule[1.1pt]
\textbf{CVE}   & \textbf{Library} & \textbf{Description}                           \\ \midrule[.9pt]
CVE-2019-1563  & OpenSSL          & Decrypt encrypted message                       \\ \hline
\rowcolor{lightblue} CVE-2017-16544 & BusyBox          & Allow executing arbitrary code                 \\ \hline
CVE-2016-6303  & OpenSSL          & Integer overflow                               \\ \hline
\rowcolor{lightblue} CVE-2016-6302  & OpenSSL          & Allows denial-of-service                       \\ \hline
CVE-2016-2842  & OpenSSL          & Allows denial-of-service                       \\ \hline
\rowcolor{lightblue} CVE-2016-2182  & OpenSSL          & Allows denial-of-service                       \\ \hline
CVE-2016-2180  & OpenSSL          & Out-of-bounds read                             \\ \hline
\rowcolor{lightblue} CVE-2016-2178  & OpenSSL          & Leak DSA private key                           \\ \hline
CVE-2016-2176  & OpenSSL          & Buffer over-read                               \\ \hline
\rowcolor{lightblue} CVE-2016-2109  & OpenSSL          & Allows denial-of-service                       \\ \hline
CVE-2016-2106  & OpenSSL          & Integer overflow                               \\ \hline
\rowcolor{lightblue} CVE-2016-2105  & OpenSSL          & Integer overflow                               \\ \hline
CVE-2016-0799  & OpenSSL          & Out-of-bounds read                             \\ \hline
\rowcolor{lightblue} CVE-2016-0798  & OpenSSL          & Allows denial-of-service                       \\ \hline
CVE-2016-0797  & OpenSSL          & NULL pointer dereference                       \\ \hline
\rowcolor{lightblue} CVE-2016-0705  & OpenSSL          & Memory corruption                              \\ \bottomrule[1.1pt]
\end{tabular}

\end{table}

Table~\ref{tab:cve_case} shows 16 vulnerabilities (CVEs) we use to search in the firmware images.
We focus on the CVEs of OpenSSL and BusyBox, as they are widely included in the firmware.
For each CVE, we compile the corresponding vulnerable functions in the specified library version and computes the vulnerable function embeddings via \sys.
As the firmware images are stripped so that we do not know with which optimizations they are compiled, we compile the vulnerable functions to both MIPS and ARM with \texttt{-O3} and rely on \sys's capability in cross-architecture and optimization function matching to match functions that are potentially compiled in different architectures and with different optimizations.
We then obtain the firmware functions that rank top-10 similar to the vulnerable function and manually verify if they are vulnerable.
We leverage \texttt{strings} command to identify the OpenSSL and BusyBox versions indicative of the corresponding vulnerabilities.
Note that such information can be stripped for other libraries so it is not a reliable approach in general.
As shown in Table~\ref{tab:cve_case}, we have confirmed all 16 CVEs in 4 firmware models developed by well-known vendors, \ie Ubiquiti, TP-Link, NETGEAR, and Linksys.
These cases demonstrate the practicality of \sys, which helps discover real-world vulnerabilities in large-scale firmware databases.
Table~\ref{tab:cve} shows the details of the 16 vulnerabilities that \sys uncover in the 4 firmware images shown in Table~\ref{tab:cve_case}.
The description of ``allow denial-of-service'' usually refers to the segmentation fault that crashes the program. 
The cause of such error can be diverse, which is not due to the other typical causes shown in the table (\eg integer overflow, buffer over-read, etc.).

\section{Related Work}
\label{sec:related}

\subsection{Binary Similarity}

\vspace{.1cm}\noindent\textbf{Traditional approaches.}
Existing static approaches often extract hand-crafted features by domain experts to match similar functions.
The features often encode the functions' syntactic characteristics. 
For example, BinDiff~\cite{bindiff} extracts the number of basic blocks and the number of function calls to determine the similarity.
Other works~\cite{myles2005k, khoo2013rendezvous, crussell2013andarwin, farhadi2014binclone} introduce more carefully-selected static features such as n-gram of instruction sequences. 
Another popular approach is to compute the structural distance between functions to determine the similarity~\cite{bindiff, bourquin2013binslayer, david2014tracelet, pewny2014leveraging, eschweiler2016discovre, david2017similarity, huang2017binsequence}.
For example, BinDiff~\cite{bindiff} performs graph matching between functions' call graphs.
TEDEM~\cite{pewny2014leveraging} matches the basic block expression trees.
BinSequence~\cite{huang2017binsequence} and Tracelet~\cite{david2014tracelet} uses the edit distance between functions' instruction sequence.
As discussed in Section~\ref{sec:intro}, both static features and structures are susceptible to obfuscations and optimizations and incur high overhead.
\sys automates learning approximate execution semantics without any manual effort, and the execution semantics is more robust to match semantically similar functions.

In addition to the static approaches, dynamic approaches such as iLine~\cite{jang2013towards}, BinHunt~\cite{ming2012ibinhunt}, iBinHunt~\cite{ming2012ibinhunt}, Revolver~\cite{kapravelos2013revolver}, Blex~\cite{egele2014blanket}, Rieck \etal\cite{rieck2011automatic}, Multi-MH~\cite{pewny2015cross}, BinGo~\cite{chandramohan2016bingo}, ESH~\cite{david2016statistical}, BinSim~\cite{ming2017binsim}, CACompare~\cite{hu2017binary}, Vulseeker-pro~\cite{gao2018vulseeker}, and Tinbergen~\cite{mckee2019software} construct hand-coded dynamic features, such as values written to stack/heap~\cite{egele2014blanket} or system calls~\cite{ming2017binsim} by executing the function to match similar functions.
These approaches can detect semantically similar (but syntactically different) functions by observing their similar execution behavior.
However, as discussed in Section~\ref{sec:intro}, these approaches can be expensive and can suffer from false positives due to the under-constrained dynamic execution traces~\cite{jiang2009automatic, ding2019asm2vec}.
By contrast, we only use these traces to learn approximate execution semantics of individual instructions and transfer the learned knowledge to match similar functions without directly comparing their dynamic traces.
Therefore, we are much more efficient and less susceptible to the imprecision introduced by these under-constrained dynamic traces.

\vspace{.1cm}\noindent\textbf{Learning-based approaches.}
Most recent learning-based works such as Genius~\cite{feng2016scalable}, Gemini~\cite{xu2017neural}, Asm2Vec~\cite{ding2019asm2vec}, SAFE~\cite{massarelli2019safe}, DeepBinDiff~\cite{duandeepbindiff} learn a function representation that is supposed to encode the function syntax and semantics in low dimensional vectors, known as function embeddings.
The embeddings are constructed by learning a neural network that takes the functions' structures (control flow graph)~\cite{feng2016scalable,xu2017neural,duandeepbindiff} or instruction sequences~\cite{ding2019asm2vec, massarelli2019safe} as input and train the model to align the function embedding distances to the similarity scores.
All existing approaches are based only on static code, which lacks the knowledge of function execution semantics.
Moreover, the learning architectures adopted in these approaches require constructing expensive graph features (attributed CFG~\cite{feng2016scalable, xu2017neural}) or limited in modeling long-range dependencies (based on Word2Vec~\cite{ding2019asm2vec, duandeepbindiff}).
By contrast, \sys learns approximate execution semantics to match functions.
Its underlying architecture is amenable to learning long-range dependencies in sequences without heavyweight feature engineering or graph construction.

\subsection{Learning Program Representations}

There has been a growing interest in learning neural program representation as embeddings from ``Big Code''~\cite{allamanis2018survey}.
The learned embedding of the code encodes the program's key properties (\ie semantics, syntax), which can be leveraged to perform many applications beyond function similarity, such as program repair~\cite{parihar2017automatic, wang2017semantics}, recovering symbol names and types~\cite{chua2017neural, patrick2020probabilistic}, code completion~\cite{raychev2014code}, decompilation~\cite{fu2019coda, katz2018using}, prefetching~\cite{shi2019learning}, and many others that we refer to Allamanis \etal\cite{allamanis2018survey} for a more thorough list.
Among these works, the most closest work to us is XDA~\cite{pei2021xda}, which also leverages the transfer learning to learn general program representations for recovering function and instruction boundaries.
However, XDA only learns from static code at the raw byte level, which lacks the understanding of execution semantics.

The core technique proposed in this paper -- learning approximate execution semantics from micro-traces -- is by no means limited to only function similarity task but can be applied to any of the above tasks.
Indeed, we plan to explore how the learned semantics in our model can transfer to other (binary) program analysis tasks in our future work.

\section{Conclusion}
\label{sec:conclusion}

We introduced \sys to match semantically similar functions based on the function execution semantics.
Our key insight is to first pretrain the ML model to explicitly learn approximate execution semantics based on the functions' mico-traces and then transfer the learned knowledge to match semantically similar functions.
Our evaluation showed that the learned approximate execution semantics drastically improves the accuracy of matching semantically similar functions -- \sys excels in matching functions across different architectures, optimizations, and obfuscations.
We plan to explore in our future work how the learned execution semantics of the code can further boost the performance of broader (binary) program analysis tasks such as decompilation.



\section*{Acknowledgment}

We thank our shepherd Lorenzo Cavallaro and the anonymous reviewers for their constructive and valuable feedback. 
This work is sponsored in part by NSF grants CCF-18-45893, CCF-18-22965, CCF-16-19123, CNS-18-42456, CNS-18-01426, CNS-16-18771, CNS-16-17670, CNS-15-64055, and CNS-15-63843; ONR grants N00014-17-1-2010, N00014-16-1-2263, and N00014-17-1-2788; an NSF CAREER award; an ARL Young Investigator (YIP) award; a Google Faculty Fellowship; a JP Morgan Faculty Research Award; a DiDi Faculty Research Award; a Google Cloud grant; a Capital One Research Grant; and an Amazon Web Services grant. 
Any opinions, findings, conclusions, or recommendations expressed herein are those of the authors, and do not necessarily reflect those of the US Government, ONR, ARL, NSF, Captital One, Google, JP Morgan, DiDi, or Amazon.



\bibliographystyle{plain}
\bibliography{paper}

\begin{appendix}

\subsection{Detailed Model Architecture}
\label{subsec:detailed_arch}

\vspace{.1cm}\noindent\textbf{Multi-head self-attention.}
Given the embeddings of all input token (see Section~\ref{subsec:pretrain_method}) in $l$-th layer, $(E_{l,1},...,E_{l,n})$, the self-attention layer updates each embedding with the following steps.
It first maps $E_{l,i}$ to three embeddings, \ie query embedding $q_i$, key embedding $k_i$, and value embedding $v_i$: 
\begin{equation*}
\label{eq:kqv}
\centering
    q_i=f_{q}(W_{q};E_{l,i}); \ k_i=f_{k}(W_{k};E_{l,i}); \ v_i=f_{v}(W_{v};E_{l,i})
\end{equation*}

Here $f_{q}$, $f_{k}$, and $f_{v}$ are affine transformation functions (\ie a fully-connected layer) parameterized by $W_{q}$, $W_{k}$, and $W_{v}$, respectively.
It then computes attention $a_{ij}$ between $E_{l,i}$ and all other embeddings $E_{l,\{j|j\neq i\}}$ by taking the dot product between the $E_{l,i}$'s query embedding $q_i$ and $E_{l,j}$'s key embedding $k_j$: $a_{ij}=q_i\cdot k_j$. 
Intuitively, attention $a$ is a square matrix, where each cell $a_{ij}$ indicates how much attention $E_{l,i}$ should pay to $E_{l,j}$ when updating itself. 
It then divides every row of $a$ by $\sqrt{d_{emb}}$ (the dimension of the embedding vectors) and scale it by softmax to ensure them sum up to 1: 
\begin{equation*}
\label{eq:scale}
\centering
    a'_{ij} = \frac{\exp(a_{ij})}{\sum^n_{j=1}\exp(a_{ij})}
\end{equation*}
The scaled attention $a'_{ij}$ will be multiplied with value embedding $v_j$ and summed up:
\begin{equation*}
\label{eq:single_head}
\centering
    E^h_{k+1,i} = \sum^n_{j=1} a'_{ij} v_j
\end{equation*}

Here $h$ in $E^h_{k+1,i}$ denote the updated embedding belong to attention head $h$. Assume we have total $H$ attention heads, the updated embeddings will finally go through an 2-layer feedforward network $f_{out}$ parameterized by $W_{out}$ with skip connections~\cite{he2016deep} to update embeddings from all heads: 

\begin{equation}
\label{eq:multi_head}
\centering
    E_{l+1,i} = f_{out}(concat(E^0_{l+1,i},...,E^H_{l+1,i});W_{out})
\end{equation}


\vspace{.1cm}\noindent\textbf{How to use the embedding.}
The learned embeddings $E_{l,i}$ after the last self-attention layer encodes the execution semantics of each instruction and the overall function.
Consider predicting masked input in pretraining.
Let layer $l$ be the $g_p$'s last self-attention layer.
The model will stack a 2-layer multi-layer perceptron (MLP) for predicting the masked codes and values: 

\begin{equation*}
    MLP(E_{l,i}) = softmax(tanh(E_{l,i}\cdot W_1)\cdot W_2), i\in\mathbb{MP}
\end{equation*}

$W_1\in \mathbb{R}^{d_{emb}\times d_{emb}}$ and $W_2\in \mathbb{R}^{d_{emb}\times |V|}$ where $|V|$ is the vocabulary size of the code token or bytes. 
As shown in Equation~\ref{eq:pretrain}, each embedding will have 9 stacked MLPs (\ie one for predicting code and the rest for predicting bytes).



\subsection{\sys Hyperparameters}
\label{subsec:hyperparm}


\vspace{.1cm}\noindent\textbf{Network architecture.} We use 12 self-attention layers with each having 8 self-attention heads. 
The embedding dimension is $d_{emb}=d_{func}=768$, which is also the embedding dimension used in bi-LSTM.
We set 3072 as the hidden layer size of MLP in the self-attention layer. 
We adopt GeLU~\cite{hendrycks2016gaussian}, known for addressing the problem of vanishing gradient, as the activation function for \sys's self-attention module. We use the hyperbolic tangent (tanh) as the activation function in finetuning MLP. 
We set the dropout rate 0.1 for pretraining do not use dropout in finetuning.

\vspace{.1cm}\noindent\textbf{Pretraining.} 
We fix the largest input length to 512 and choose the effective batch size for both pretraining and finetuning as 512.
As 512 batch size with each sample of length 512 is too large to fit in our GPU memory (11 GB), we aggregate the gradient every 64 batches and then updates the weight parameter.
This setting results in the actual batch size as 8 ($64\times8=512$). 
We pick learning rate $5\times10^{-4}$ for pretraining.
Instead of starting with the chosen learning rate at first epoch, we follow the common practice of using small warmup learning rate at first epoch. 
We use $10^{-7}$ as the initial warmup learning rate and gradually increases it until reaching the actual learning rate ($5\times10^{-4}$) after first epoch.
We use use Adam optimizer, with $\beta_1=0.9$, $\beta_2=0.98$, $\epsilon=10^{-6}$, and weight decay $10^{-2}$. 

\subsection{More Experiments}
\label{subsec:more_exp}

\vspace{.1cm}\noindent\textbf{Cross-project generalizability.} 
While we strictly separate the functions for pretraining, finetuning, and testing, the functions of finetuning and testing can come from the same project (but strictly different functions).
Therefore, in this section, we further separate the functions in finetuning and testing by extracting them from \emph{different projects}.
For example, we can finetune the model on Coreutils while testing on OpenSSL.
Specifically, we select 3 projects, \ie Binutils, Coreutils, and OpenSSL, which have the largest number of functions, and evaluate how \sys performs.
We allow the functions to come from different architectures, optimizations, and obfuscations, and follow the same setup as described in Section~\ref{sec:impl}.

\begin{table}[!t]

\footnotesize
\setlength{\tabcolsep}{7pt}
\centering
\renewcommand{\arraystretch}{1.1}
\setlength\aboverulesep{0.4pt}
\setlength\belowrulesep{0.4pt}

\caption{\sys results (in AUC score) on function pairs with training and testing functions extracted from different projects.
}
\label{tab:cross-dataset}

\begin{tabular}{r|c|c|c}
\toprule[1.1pt]
\backslashbox{Train}{Test} & \textbf{Coreutils} & \textbf{Binutils} & \textbf{OpenSSL} \\ \midrule[.9pt]
\rowcolor{lightblue} \textbf{Coreutils} & 0.947 & 0.945 & 0.940 \\ \hline
\textbf{Binutils} & 0.945 & 0.945 & 0.944 \\ \hline
\rowcolor{lightblue} \textbf{OpenSSL} & 0.936 & 0.939 & 0.956 \\ \bottomrule[1.1pt]
\end{tabular}

\end{table}

Table~\ref{tab:cross-dataset} shows that \sys's AUC score does not drop dramatically ($<2$\%) when the functions are coming from different projects when compared to coming from same projects (the diagonal). 
This observation indicates \sys generalizes to unseen function pairs in an entirely different dataset.
Note that the functions in each function pair can come from arbitrary architectures, optimizations, and obfuscations (last column in Table~\ref{tab:result}).
As we have shown in Section~\ref{subsec:rq2}, the numbers achieved by \sys in Table~\ref{tab:cross-dataset} even outperforms the existing baselines when their functions can only come from different architectures and within only the same projects.

\vspace{.1cm}\noindent\textbf{Effectiveness of bi-LSTM encoding.}
As described in Section~\ref{subsec:input_repr}, we treat the numeric values as an 8-byte sequence and use bi-LSTM to combine them into a single representation (embedding).
The structure of bi-LSTM is known to capture the potential dependencies between different bytes (with different significance) in the byte-sequence.
Indeed, besides bi-LSTM, there are other possible differentiable modules such as multi-layer perceptron (MLP) or simple summation can also combine the input bytes.
Therefore, we study the performance of \sys in predicting masked tokens when we vary the modules for byte-sequence combination.
Specifically, we follow the same setting described above by selecting a random 10,000 function binaries as the testing set and evaluate the testing PPL achieved by pretraining \sys.

Figure~\ref{fig:pretrain-loss} shows the validation PPL in 10 epochs of pretraining \sys based on (1) bi-LSTM, (2) 2-layer (with 1024 hidden size) MLP, and (3) simple summation (SUM), to combine the byte-sequence.
We can observe that bi-LSTM is obviously better than other two, achieving the lowest PPL.
This indicates that bi-LSTM helps \sys the most in terms of learning approximate execution semantics.

\vspace{.1cm}\noindent\textbf{Vulnerability search performance.}
We quantify the accuracy of \sys in searching vulnerable functions in the firmware images and compare it to that of SAFE.
As SAFE does not work for MIPS, we study how it performs on NETGEAR R7000 model, the only model that runs on ARM architecture from Table~\ref{tab:cve_case}.
Specifically, we compile OpenSSL to ARM and x64 with \texttt{O3}, and feed both our compiled and firmware's function binaries to \sys and SAFE to compute embeddings.
Based on the function embeddings, we search the compiled OpenSSL functions in the NETGEAR R7000's embedded OpenSSL libraries, and test their top-1/3/5/10 errors.
For example, the top-10 error measures when the query function does not appear in the top-10 most similar functions in the firmware.

\begin{figure}[!t]
\centering

\subfloat[Same architecture]{
\includegraphics[width=0.47\linewidth]{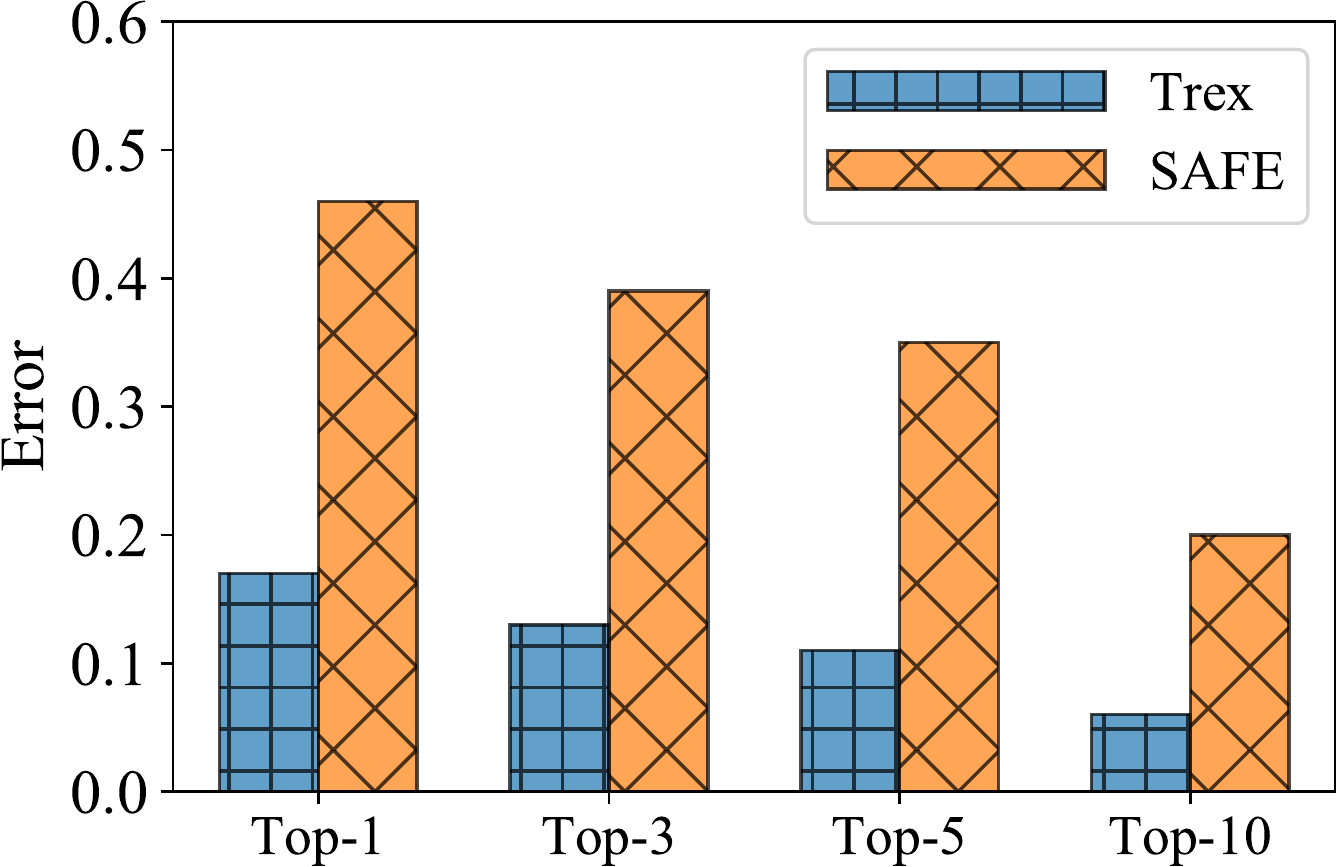}
\label{subfig:same}}
\subfloat[Cross architecture]{
\includegraphics[width=0.47\linewidth]{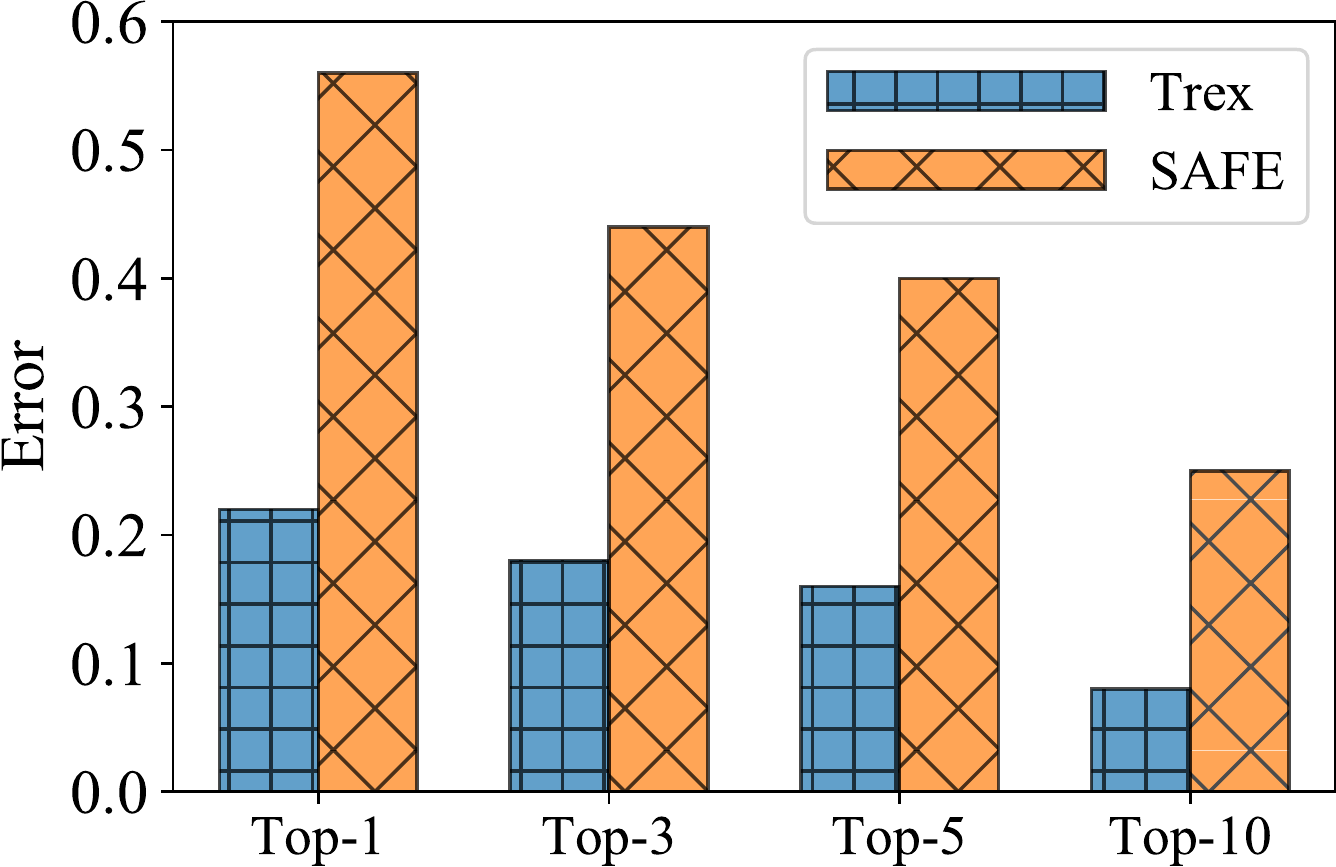}
\label{subfig:cross}}

\caption{Top-1/3/5/10 error of \sys and SAFE in searching functions in firmware. (a) The query functions and the firmware are from same architecture (ARM). (b) The query functions are from x64 but the firmware are from ARM.}
\label{fig:vuln_error}
\end{figure}

Figure~\ref{fig:vuln_error} shows that \sys consistently outperforms SAFE, achieving 24.3\% lower error rate on average.
Moreover, as shown in Figure~\ref{subfig:cross}, when the query functions are from x64 (the firmware binaries are from ARM), \sys outperforms SAFE by a greater margin, achieving 25.3\% lower error rate in cross-architecture search.

\begin{table*}[!t]

\footnotesize
\setlength{\tabcolsep}{1.5pt}
\centering
\renewcommand{\arraystretch}{1}
\setlength\aboverulesep{0.4pt}
\setlength\belowrulesep{0.4pt}

\caption{Details of the real-world 180 firmware images we collected from DD-WRT. 
}
\label{tab:firmware}

\begin{tabular}{l|l|l|l?{1.1pt}l|l|l|l?{1.1pt}l|l|l|l}
\toprule[1.1pt]
\textbf{Vendor} & \textbf{Model} & \textbf{ISA} & \textbf{Year} & \textbf{Vendor} & \textbf{Model} & \textbf{ISA} & \textbf{Year} & \textbf{Vendor} & \textbf{Model} & \textbf{ISA} & \textbf{Year}  \\ \midrule[.9pt]

\rowcolor{lightblue}	8devices	&Carambola 2	&MIPS	&2020	&8devices	&Carambola 2 8MB	&MIPS	&2018	&8devices	&Lima	&MIPS	&2020	 \\ \hline
	Actiontec	&MI424WR	&ARM	&2019	&Alfa	&AIP-W502U	&MIPS	&2020	&Alfa	&SOLO48	&MIPS	&2020	 \\ \hline
\rowcolor{lightblue}	Belkin	&F5D8235-4	&MIPS	&2020	&Buffalo	&BHR-4GRV	&MIPS	&2020	&Buffalo	&WBMR-HP-G300H	&MIPS	&2020	 \\ \hline
	Buffalo	&WHR-1166D	&MIPS	&2020	&Buffalo	&WHR-300HP2	&MIPS	&2018	&Buffalo	&WHR-600D	&MIPS	&2018	 \\ \hline
\rowcolor{lightblue}	Buffalo	&WXR-1900DHP	&ARM	&2020	&Buffalo	&WZR-1166DHP	&ARM	&2020	&Buffalo	&WZR-600DHP2	&ARM	&2020	 \\ \hline
	Buffalo	&WZR-900DHP	&ARM	&2020	&Buffalo	&WZR-HP-G450H	&MIPS	&2020	&Comfast	&CF-E325N	&MIPS	&2020	 \\ \hline
\rowcolor{lightblue}	Comfast	&CF-E355AC	&MIPS	&2020	&Comfast	&CF-E380AC	&MIPS	&2020	&Comfast	&CF-WR650AC	&MIPS	&2020	 \\ \hline
	Compex	&WP546	&MIPS	&2020	&Compex	&WPE72	&MIPS	&2020	&D-Link	&DAP-2230	&MIPS	&2020	 \\ \hline
\rowcolor{lightblue}	D-Link	&DAP-2330	&MIPS	&2020	&D-Link	&DAP-2660	&MIPS	&2020	&D-Link	&DAP-3320	&MIPS	&2020	 \\ \hline
	D-Link	&DAP-3662	&MIPS	&2020	&D-Link	&DHP-1565 A1	&MIPS	&2020	&D-Link	&DIR-825 C1	&MIPS	&2020	 \\ \hline
\rowcolor{lightblue}	D-Link	&DIR-825 Rev.B	&MIPS	&2020	&D-Link	&DIR-835 A1	&MIPS	&2020	&D-Link	&DIR-859	&MIPS	&2020	 \\ \hline
	D-Link	&DIR-860L A1	&ARM	&2020	&D-Link	&DIR-860L B1	&MIPS	&2020	&D-Link	&DIR-862	&MIPS	&2020	 \\ \hline
\rowcolor{lightblue}	D-Link	&DIR-866	&MIPS	&2020	&D-Link	&DIR-868l Rev.A	&ARM	&2020	&D-Link	&DIR-868l Rev.B	&ARM	&2020	 \\ \hline
	D-Link	&DIR-868l Rev.C	&ARM	&2020	&D-Link	&DIR-869	&MIPS	&2020	&D-Link	&DIR-878 A1	&MIPS	&2020	 \\ \hline
\rowcolor{lightblue}	D-Link	&DIR-880l	&ARM	&2020	&D-Link	&DIR-882 A1	&MIPS	&2020	&D-Link	&DIR-885l	&ARM	&2020	 \\ \hline
	D-Link	&DIR-890l	&ARM	&2020	&D-Link	&DIR632A	&MIPS	&2020	&GL.iNet	&AR150	&MIPS	&2020	 \\ \hline
\rowcolor{lightblue}	Gateworks	&GW2382	&ARM	&2018	&Gateworks	&GW2388 16M	&ARM	&2018	&Gateworks	&GW2391	&ARM	&2018	 \\ \hline
	Gateworks	&Laguna GW2382	&ARM	&2020	&Gateworks	&Laguna GW2388 32M	&ARM	&2020	&Gateworks	&Laguna GW2391	&ARM	&2020	 \\ \hline
\rowcolor{lightblue}	Gigaset	&SX763	&MIPS	&2020	&JJPlus	&JA76PF	&MIPS	&2020	&Linksys	&E1700	&MIPS	&2020	 \\ \hline
	Linksys	&E2100L	&MIPS	&2020	&Linksys	&EA6350	&ARM	&2020	&Linksys	&EA6400	&ARM	&2020	 \\ \hline
\rowcolor{lightblue}	Linksys	&EA6500 v2	&ARM	&2020	&Linksys	&EA6700	&ARM	&2020	&Linksys	&EA6900	&ARM	&2020	 \\ \hline
	Linksys	&EA8500	&ARM	&2020	&Linksys	&RE7000	&MIPS	&2020	&Linksys	&WRT610N v1.0	&MIPS	&2020	 \\ \hline
\rowcolor{lightblue}	NETGEAR	&AC1450	&ARM	&2020	&NETGEAR	&EX6200	&ARM	&2018	&NETGEAR	&R6250	&MIPS	&2020	 \\ \hline
	NETGEAR	&R6300v2	&ARM	&2020	&NETGEAR	&R6400	&ARM	&2020	&NETGEAR	&R6700	&ARM	&2020	 \\ \hline
\rowcolor{lightblue}	NETGEAR	&R6700v3	&ARM	&2020	&NETGEAR	&R7000	&ARM	&2020	&NETGEAR	&R7000P	&ARM	&2020	 \\ \hline
	NETGEAR	&R7500v1	&ARM	&2020	&NETGEAR	&R7500v2	&ARM	&2020	&NETGEAR	&R7800	&ARM	&2020	 \\ \hline
\rowcolor{lightblue}	NETGEAR	&R8500	&ARM	&2020	&NETGEAR	&R8900	&ARM	&2020	&NETGEAR	&R9000	&ARM	&2020	 \\ \hline
	NETGEAR	&WG302v2	&ARM	&2020	&NETGEAR	&WNDR3700v4	&MIPS	&2020	&NETGEAR	&WNDR4300	&MIPS	&2020	 \\ \hline
\rowcolor{lightblue}	NETGEAR	&WNDR4500	&MIPS	&2020	&NETGEAR	&WNDR4500v2	&MIPS	&2020	&NETGEAR	&XR450	&ARM	&2020	 \\ \hline
	NETGEAR	&XR500	&ARM	&2020	&NETGEAR	&XR700	&ARM	&2020	&Pronghorn	&SBC	&ARM	&2020	 \\ \hline
\rowcolor{lightblue}	Senao	&ECB3500	&MIPS	&2020	&Senao	&ECB9750	&MIPS	&2019	&Senao	&EOC1650	&MIPS	&2020	 \\ \hline
	Senao	&EOC5510	&MIPS	&2020	&Senao	&EOC5610	&MIPS	&2020	&Senao	&EOC5611	&MIPS	&2020	 \\ \hline
\rowcolor{lightblue}	Senao	&NOP8670	&ARM	&2020	&TP-Link	&Archer A7 V5	&MIPS	&2020	&TP-Link	&Archer C1900	&ARM	&2020	 \\ \hline
	TP-Link	&Archer C25 V1	&MIPS	&2020	&TP-Link	&Archer C5 V1	&MIPS	&2020	&TP-Link	&Archer C7 V1	&MIPS	&2020	 \\ \hline
\rowcolor{lightblue}	TP-Link	&Archer C7 V2	&MIPS	&2020	&TP-Link	&Archer C7 V3	&MIPS	&2020	&TP-Link	&Archer C7 V4	&MIPS	&2020	 \\ \hline
	TP-Link	&Archer C8 V1	&ARM	&2020	&TP-Link	&Archer C9 V1	&ARM	&2020	&TP-Link	&Archer C9 V2	&ARM	&2020	 \\ \hline
\rowcolor{lightblue}	TP-Link	&Archer C9 V3	&ARM	&2020	&TP-Link	&Deco-M4	&MIPS	&2020	&TP-Link	&TL-WDR3600 V1	&MIPS	&2020	 \\ \hline
	TP-Link	&TL-WDR4300 V1	&MIPS	&2020	&TP-Link	&TL-WDR4310 V1	&MIPS	&2020	&TP-Link	&TL-WDR4900 V2	&MIPS	&2020	 \\ \hline
\rowcolor{lightblue}	TP-Link	&TL-WR1043N V5	&MIPS	&2020	&TP-Link	&TL-WR1043ND	&MIPS	&2020	&TP-Link	&TL-WR1043ND V2	&MIPS	&2020	 \\ \hline
	TP-Link	&TL-WR1043ND V4	&MIPS	&2020	&TP-Link	&TL-WR2543ND	&MIPS	&2020	&TP-Link	&TL-WR710N V1	&MIPS	&2020	 \\ \hline
\rowcolor{lightblue}	TP-Link	&TL-WR710N V2.1.0	&MIPS	&2020	&TP-Link	&TL-WR810N V1	&MIPS	&2020	&TP-Link	&TL-WR810N V2	&MIPS	&2020	 \\ \hline
	TP-Link	&TL-WR842N V1	&MIPS	&2020	&TP-Link	&TL-WR842N V2	&MIPS	&2020	&TRENDnet	&TEW-811DRU	&ARM	&2018	 \\ \hline
\rowcolor{lightblue}	TRENDnet	&TEW-812DRU V2	&ARM	&2018	&TRENDnet	&TEW-818DRU	&ARM	&2020	&TRENDnet	&TEW-828DRU	&ARM	&2020	 \\ \hline
	Ubiquiti	&AirGrid M2	&MIPS	&2020	&Ubiquiti	&AirGrid M5	&MIPS	&2020	&Ubiquiti	&AirGrid-M5-XW	&MIPS	&2020	 \\ \hline
\rowcolor{lightblue}	Ubiquiti	&AirRouter	&MIPS	&2020	&Ubiquiti	&AirRouter-HP	&MIPS	&2020	&Ubiquiti	&AirWire	&MIPS	&2020	 \\ \hline
	Ubiquiti	&BulletM2 HP	&MIPS	&2020	&Ubiquiti	&BulletM5 HP	&MIPS	&2020	&Ubiquiti	&LS SR71A	&MIPS	&2020	 \\ \hline
\rowcolor{lightblue}	Ubiquiti	&NanoBeam AC	&MIPS	&2020	&Ubiquiti	&NanoBeam M2 XW	&MIPS	&2020	&Ubiquiti	&NanoBeam M5 XW	&MIPS	&2020	 \\ \hline
	Ubiquiti	&NanoBridge M2	&MIPS	&2020	&Ubiquiti	&NanoBridge M2 XW	&MIPS	&2020	&Ubiquiti	&NanoBridge M3	&MIPS	&2020	 \\ \hline
\rowcolor{lightblue}	Ubiquiti	&NanoBridge M365	&MIPS	&2020	&Ubiquiti	&NanoBridge M5 XW	&MIPS	&2020	&Ubiquiti	&NanoBridge M900	&MIPS	&2020	 \\ \hline
	Ubiquiti	&NanoStation M2	&MIPS	&2020	&Ubiquiti	&NanoStation M3	&MIPS	&2020	&Ubiquiti	&NanoStation M365	&MIPS	&2020	 \\ \hline
\rowcolor{lightblue}	Ubiquiti	&Pico M5	&MIPS	&2020	&Ubiquiti	&Power AP N	&MIPS	&2020	&Ubiquiti	&PowerBeam M5 M400 XW	&MIPS	&2020	 \\ \hline
	Ubiquiti	&PowerBridge M10	&MIPS	&2020	&Ubiquiti	&PowerBridge M5	&MIPS	&2020	&Ubiquiti	&Rocket M2 Titanium XW	&MIPS	&2020	 \\ \hline
\rowcolor{lightblue}	Ubiquiti	&Rocket M2 XW	&MIPS	&2020	&Ubiquiti	&Rocket M5 Titanium XW	&MIPS	&2020	&Ubiquiti	&Rocket M5 X3 XW	&MIPS	&2020	 \\ \hline
	Ubiquiti	&Rocket M5 XW	&MIPS	&2020	&Ubiquiti	&RocketM2	&MIPS	&2020	&Ubiquiti	&RocketM3	&MIPS	&2020	 \\ \hline
\rowcolor{lightblue}	Ubiquiti	&RocketM365	&MIPS	&2020	&Ubiquiti	&RocketM5	&MIPS	&2020	&Ubiquiti	&RocketM900	&MIPS	&2020	 \\ \hline
	Ubiquiti	&RouterStation	&MIPS	&2020	&Ubiquiti	&RouterStation Pro	&MIPS	&2020	&Ubiquiti	&sunMax	&MIPS	&2020	 \\ \hline
\rowcolor{lightblue}	Ubiquiti	&UAP-AC-MESH	&MIPS	&2020	&Ubiquiti	&UAP-AC-PRO	&MIPS	&2020	&Ubiquiti	&UAP-LR	&MIPS	&2020	 \\ \hline
	Ubiquiti	&UAP-LR-v2	&MIPS	&2020	&Ubiquiti	&UAP-v2	&MIPS	&2020	&Ubiquiti	&locoM2	&MIPS	&2020	 \\ \hline
\rowcolor{lightblue}	Ubiquiti	&locoM2 XW	&MIPS	&2020	&Ubiquiti	&locoM5	&MIPS	&2020	&Ubiquiti	&locoM5 XW	&MIPS	&2020	 \\ \hline
	Ubiquiti	&locoM900	&MIPS	&2020	&WiliGear	&WBD-500	&MIPS	&2020	&YunCore	&XD3200	&MIPS	&2020	 \\ \bottomrule[1.1pt]

\end{tabular}

\end{table*}

\subsection{Probing Learned Execution Semantics}
\label{subsec:probe_case}

As discussed in Section~\ref{sec:overview}, training the model to predict masked micro-trace codes and values compels the model to learn execution semantics by concrete examples.
It thus automates the extraction of the code's dynamic features without manual effort.
In this section, we study concrete code examples showing the potential hint in the code that the model likely leverages to predict the masked part. 

\begin{figure}[!t]
\centering

\includegraphics[width=\linewidth]{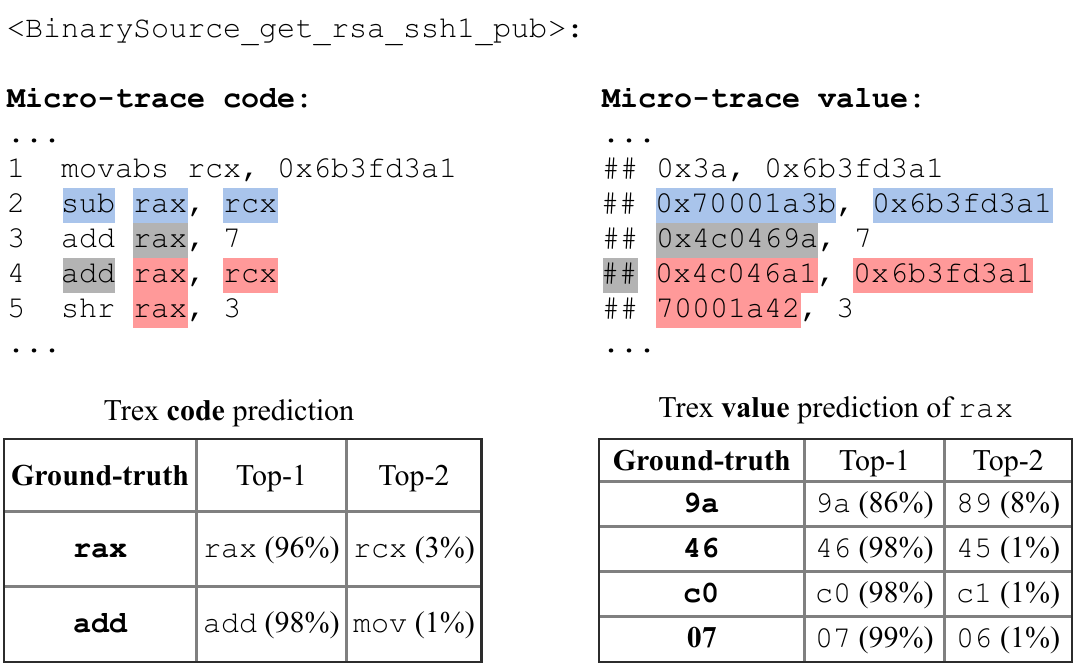}

\caption{The partial micro-trace code and value sequence from function \texttt{BinarySource\_get\_rsa\_ssh1\_pub} in PuTTy-0.74 compiled to x64 with \texttt{O0}. We mask the register \colorbox{lightgray}{\texttt{rax}} at line 3 and the opcode \colorbox{lightgray}{\texttt{add}} at line 4. We also mask their corresponding micro-trace values (where opcode has only dummy values (8 \texttt{\#\#}s) as described in Section~\ref{subsec:input_repr}). 
We highlight the contextual hint in \colorbox{myblue}{blue} that \sys leverages to predict the masked \colorbox{lightgray}{\texttt{rax}} and \colorbox{lightred}{red} that \sys uses to predict the masked \colorbox{lightgray}{\texttt{add}}.}
\label{fig:arithmetic}
\end{figure}

\vspace{.1cm}\noindent\textbf{Predicting arithmetic instructions.}
Consider the example in Figure~\ref{fig:arithmetic}, which shows the function's micro-trace (\ie the dynamic value and assembly code). 
For ease of exposition, the format of micro-trace values (\eg the value of opcode and constants) are not exactly the same as the actual format we handle, as described in Section~\ref{subsec:input_repr}.
We mask the register \texttt{rax} and opcode \texttt{add} and their corresponding values in a sequence of arithmetic instructions.

To correctly predict the masked value \texttt{0x4c0469a} of \texttt{rax}, the model should understand the approximate execution semantics of \texttt{sub} in line 2, which subtracts \texttt{rax} with \texttt{rcx}.
It may also observe the value of \texttt{rax} at line 4, which exactly equals the result of adding \texttt{rax} with 7 at line 3. 
Therefore, we see the model predicts \texttt{rax} with 96\% confidence and \texttt{rcx} with only 3\%.
It also predicts the value of \texttt{rax} correctly (right table).
To correctly predict \texttt{add} at line 4, the model should observe that the \texttt{rax} at line 5 has the same exact value with the result of \texttt{rax+rcx} at line 4, so it predicts the opcode at line 4 to be \texttt{add}.
As shown in Figure~\ref{fig:arithmetic}, our pretrained model predicts \texttt{add} with 98\% confidence and \texttt{mov} with only 1\%, which implies that the model approximately understands the execution semantics of \texttt{add} and \texttt{mov} so it predicts that \texttt{add} is much more likely. 

\vspace{.1cm}\noindent\textbf{Predicting stack operations.}
Consider the example in Figure~\ref{fig:stack}. 
We mask the register and constant of the instruction in function epilogue -- it increments the stack pointer \texttt{rsp} by \texttt{0x20} to deallocate the local variable stored on stack.
To correctly predict the masked \texttt{rsp} and its value \texttt{0x8b4a3f}, the model should observe the \texttt{rsp} is decremented (due to opcode \texttt{sub}) by \texttt{0x20} from \texttt{0x8b4a3f} at line 3, which is part of the function prologue.
Therefore, the model should understand the execution semantics of \texttt{sub} and basic syntax of function prologue and epilogue.

To predict the masked constant \texttt{0x20}, the model should notice that line 3 decrements the stack pointer by \texttt{0x20}, which is the size of local variables.
As the model predicts \texttt{0x20} with 99\% confidence, this implies the model likely learns patterns of function prologue and epilogue, including the context such as \texttt{push ebp}, \texttt{pop ebp}, etc.
This observation indicates that our pretraining task assists in learning common function idiom (function calling convention) beyond execution semantics of individual instructions.
Learning such knowledge can potentially help other tasks beyond function similarity, such as identifying function boundaries~\cite{pei2021xda}.

\begin{figure}[!t]
\centering

\includegraphics[width=.95\linewidth]{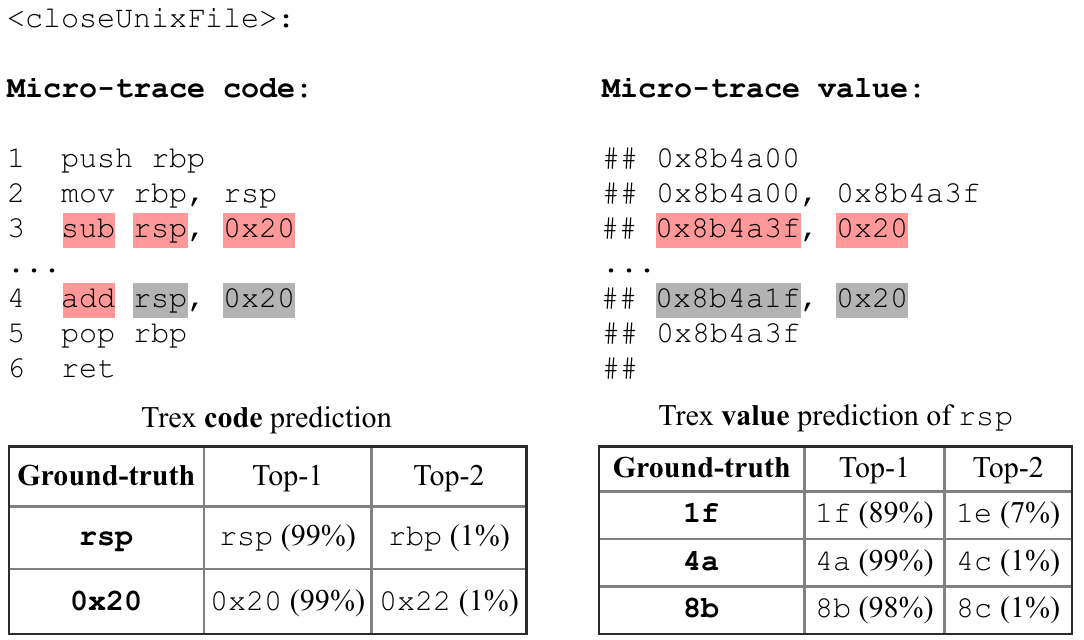}

\caption{The partial code and micro-trace from function \texttt{closeUnixFile} in SQLite-3.34.0 compiled to x64 with \texttt{O0}.
We mask the register \colorbox{lightgray}{\texttt{rsp}} and constant \colorbox{lightgray}{\texttt{0x20}} at line 4. We also mask their corresponding micro-trace values.
We highlight the contextual hint in \colorbox{lightred}{red} that \sys leverages to predict the masked \colorbox{lightgray}{\texttt{rsp}} and \colorbox{lightgray}{\texttt{0x20}}.}
\label{fig:stack}
\end{figure}

\end{appendix}

\end{document}